\documentclass[a4paper,11pt]{article}

\usepackage{bbm}
\usepackage{latexsym}
\usepackage{graphicx}
\usepackage{psfrag}

\newcommand{\be}{\begin{equation}}
\newcommand{\ee}{\end{equation}}
\newcommand{\bea}{\begin{eqnarray}}
\newcommand{\eea}{\end{eqnarray}}
\newcommand{\bean}{\begin{eqnarray*}}
\newcommand{\eean}{\end{eqnarray*}}

\renewcommand{\b}{\langle}
\renewcommand{\k}{\rangle}

\renewcommand{\i}{{\rm i}}
\newcommand{\e}{{\rm e}}

\renewcommand{\d}{{\rm d}}
\renewcommand{\c}[1]{{\cal #1}}

\newcommand{\pa}{\partial}

\renewcommand{\v}[1]{\vec{#1}} 

\newcommand{\ts}{\textstyle}

\newcommand{\ds}{\displaystyle}

\newcommand{\ssize}{\scriptsize}

\newcommand{\diff}[2]{\frac{{\rm d}#1}{{\rm d}#2}}
\newcommand{\fdiff}[2]{\frac{{\rm \delta}#1}{{\rm \delta}#2}}
\newcommand{\pdiff}[2]{\frac{\partial #1}{\partial #2}}

\newcommand{\bZ}{\mathbbm{Z}}
\newcommand{\bN}{\mathbbm{N}}

\newcommand{\bR}{\mathbbm{R}}

\renewcommand{\rm}{{}}  

\renewcommand{\H}{\c{H}}
\renewcommand{\L}{\c{L}}

\newcommand{\eq}[1]{(\ref{#1})}

\newcommand{\fig}[1]{Fig.\ \ref{#1}}

\newcommand{\pic}[4]
{
  \begin{figure}
  \begin{center}
  \includegraphics[height=#3]{#4}
  \end{center}
  \caption{\label{#1} #2}
  \end{figure}
}

\newcommand{\qed}{\hspace{2em}$\Box$\vspace{1em}}

\newcommand{\ti}{t_{\rm i}}
\newcommand{\tf}{t_{\rm f}}
\newcommand{\varphii}{\varphi_{\rm i}}
\newcommand{\varphif}{\varphi_{\rm f}}
\newcommand{\pif}{\pi_{\rm f}}
\newcommand{\varphim}{\varphi_{\rm m}}
\newcommand{\varphia}{\varphi_{\rm a}}
\newcommand{\varphib}{\varphi_{\rm b}}
\newcommand{\varphic}{\varphi_{\rm c}}

\newcommand{\naSig}{\nabla_{\!\Sigma}}

\begin{document}
\thispagestyle{empty}
\hfill
\parbox[t]{3.6cm}{
hep-th/0310246 \\
AEI-2003-088}
\vspace{2cm}

\begin{center}
{\bf\Large Generalized Schr\"odinger equation in Euclidean field theory}\\[4mm] {Florian Conrady$^{1,2}$ and Carlo Rovelli$^{1,3}$\\[2mm]
$^1$\small\it Dipartimento di Fisica, Universit\`a di Roma ``La
 Sapienza'', I-00185 Rome, Italy\\
$^2$\small\it Max-Planck-Institut f\"{u}r Gravitationsphysik, 
Albert-Einstein-Institut, D-14476 Golm, Germany\\
$^3$\small\it Centre de Physique Th\'eorique de Luminy, CNRS, 
F-13288, France}
\end{center}
\vskip.5cm

\begin{abstract}
\noindent We investigate the idea of a ``general boundary"
formulation of quantum field theory in the context of the Euclidean
free scalar field. We propose a precise definition for
an evolution kernel that propagates the field through arbitrary
spacetime regions. We show that this kernel satisfies an evolution
equation which governs its dependence on deformations of the boundary surface and generalizes the ordinary (Euclidean) Schr\"odinger equation. We also derive the classical counterpart of this equation, which is a Hamilton-Jacobi equation for general boundary surfaces.
\end{abstract}
\vskip1cm

\section{Introduction}
\label{Introduction}

In quantum field theory (QFT) on Minkowski space, we can use the
Schr\"odinger picture and have states associated to flat spacelike
(hyper-)surfaces.  The transition amplitude between an initial state
and a final state is obtained by acting with the unitary evolution
operator on the former and taking the inner product with the latter. 
The possibility of a Schr\"odinger picture has also been considered in QFT on curved spacetime \cite{Helfer}.  In this case, states live on arbitrary
spacelike Cauchy surfaces forming a foliation of spacetime.  Evolution
along these surfaces is non-unitary in general, as it does not
correspond to a symmetry of the metric.  In background independent
quantum gravity, on the other hand, there is no fixed spacetime
geometry; in this case, states live on arbitrary Cauchy surfaces and
the requirement that the surface is spacelike is encoded in the state
itself, which represents a quantum state of a spacelike geometry (see
e.g.\ \cite{Thiemann,Rovelli}).

In all these cases, transition amplitudes are calculated for boundary
states (i.e.\ an initial and final state) defined on spacelike
boundaries.  Recently, Oeckl has suggested that it could be possible
to relax this restriction to {\it spacelike} boundaries in QFT
\cite{cat,Oeckl}.  Oeckl offers heuristic arguments which suggest that
transition amplitudes can be associated to a wider class of
boundaries, as we do in topological quantum field theory
\cite{Atiyah}. These more general boundaries may include hypersurfaces
which are partially timelike, that enclose a finite region of
spacetime, or disjoint unions of such sets.  This would imply, for
instance, that in theories like QED or QCD, we could associate quantum
``states" to a hypersphere, a hypercube or more exotic surfaces, and
assign probability amplitudes to them.  Similar suggestions were made
in \cite{Rovelli}, with different motivations.
 
This ``general boundary" approach to QFT could be interesting for
several reasons. Firstly, finite closed boundaries represent the way
real experiments are set up more directly than constant-time surfaces.
A realistic experiment is confined to a finite region of spacetime.  In
particle colliders, for instance, the interaction region is enclosed
by a finite outer region where state preparation and measurement take
place. As sketched in \fig{experiment},  the walls and openings of a particle detector trace out a hypercube in spacetime. 
\psfrag{t}{$t$}\psfrag{x}{$x_1$}\psfrag{y}{$x_2$}\pic{experiment}{Spacetime
diagram of particle scattering.}{4cm}{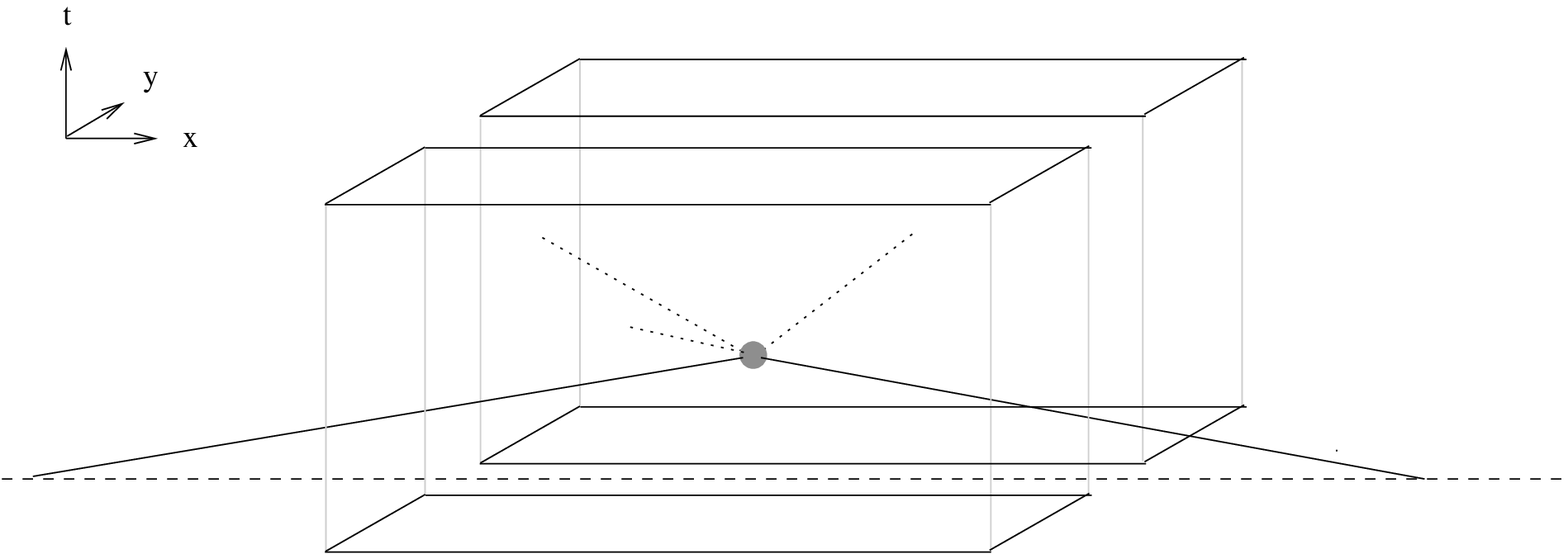} A ``state" on
the hypercube's surface would represent both incoming beams and jets
of outgoing particles in a completely local fashion, without making
any reference to inaccessible infinitely distant regions.

Secondly, in a quantum theory of gravity closed boundaries may provide a
way to define scattering amplitudes, and help in solving the traditional
interpretational difficulties of background independent QFT. This idea
has been recently studied in \cite{CDORT}, where it has also been used
to propose an explicit way for computing the Minkowski vacuum state
from a spinfoam model \cite{Baez}-\cite{Oriti}. In a background independent theory the conventional spacelike states do not impose any constraint on the proper time lapsed between the initial and final states.  As a result, the transition amplitude stems from a superposition of processes whose duration may range from
microscopic to cosmic time scales.  Fixing a timelike boundary can
control the time lapsed during the experiment.

Furthermore, in spinfoam approaches, the introduction of general boundaries might open up the possibility of quantizing 3-geometries along time-like surfaces
and clarify the physical meaning of Lorentzian spinfoams.

Finally, a general boundary formulation could give us a broader
perspective on QFT: it would stress geometrical aspects of QFT by no
longer singling out a special subclass of surfaces, and may shed
some light on the holographic principle, which states that the
complete information about a spacetime region can be encoded in its
boundary.

As noted by Oeckl, a heuristic idea for adapting QFT to general
boundaries is provided by Feynman's sum-over-paths-picture.  Given an
arbitrary spacetime region $V$, bounded by a 3d hypersurface, the
Feynman path integral over the spacetime region $V$, with fixed
boundary value $\varphi$ of the field, defines a functional
$W[\varphi,V]$.  This functional can be seen as a generalized
evolution kernel, or a generalized field propagator.  The path
integral is therefore a natural starting point for developing a
general boundary formalism.

The path to make these ideas precise is long.  There are two types of
problems. Firstly, the probabilistic interpretation of quantum theory
and QFT must be adapted to this more general case. The physical
meaning of states at fixed time and their relation to physical
measurements are well established; the extension to arbitrary
boundaries is probably doable, but far from obvious. It requires us
to treat quantum state preparation and quantum measurement on the same
ground, and to give a precise interpretation to the general
probability amplitudes.  Some steps in this direction can be found in
\cite{Oeckl} and \cite{Rovelli}. 

Secondly, the mathematical apparatus of QFT, i.e.\ the path integral and operator formalism, needs to be extended to general spacetime regions. On a formal level, such a generalization appears natural for path integrals, but it is far from clear that it can be given a concrete and well-defined meaning.

In this paper we focus on the second of these issues: the definition
of the field theoretical functional integral over an arbitrary region,
and its relation to operator equations.  We start to address the problem by considering the simplest system: Euclidean free scalar field theory.  In this context, we propose an exact definition for the propagator kernel
$W[\varphi,V]$, based on limits of lattice path integrals.  Under a number of
assumptions, we can show that the propagator satisfies a generalized
Schr\"odinger equation, of the Tomonaga-Schwinger kind
\cite{Tomonaga,Schwinger}. The equation governs the way the propagator changes
under infinitesimal deformations of $V$.  It reduces to the ordinary
Schr\"odinger equation in the case in which a boundary surface of $V$
is a constant-time surface and the deformation is a global shift in
time. 

With this result, we provide a first step towards constructing an operator formalism for general boundaries. The derivation can be seen as a higher-dimensional generalization of Feynman's path integral derivation of the Schr\"odinger equation for a single particle \cite{Feynman}.  The main assumption we need is the existence of a rotationally invariant continuum limit. 

We also derive the classical counterpart of the evolution equation: a
generalized version of the Euclidean Hamilton-Jacobi equation.  At
present, we have no prescription for Wick rotation, so we cannot give
any Lorentzian form for the propagator or the evolution equation. Hints in this direction were given in \cite{CDORT}. 

If one continues along this line, the ultimate goal would be to construct a full general boundary formalism for background dependent QFTs, which incorporates Wick rotation, interactions and renormalization. While of interest in itself, such a project could be also viewed as a testing ground for the general boundary method, which would prepare us for applying it in the more difficult context of background free QFT: there, as indicated before, the use of generalized boundary conditions may not only be helpful, but also essential for understanding the theory.

Our technique for deriving the evolution equation could be of interest in view of the attempts to relate canonical and path integral formulations of quantum gravity, i.e.\ when deriving the Wheeler-DeWitt equation from a concrete realization of a sum over geometries. (For existing results on this problem, see e.g.\ \cite{Halliwell}.) 

The paper is organized as follows.  In section
\ref{General_Boundary_Approach}, we present some of the heuristic
considerations about state functionals on general boundaries, and
their associated evolution kernel.  Section
\ref{Generalized_Hamilton_Jacobi_Equation} deals with the classical
case: we present two derivations of the generalized Hamilton-Jacobi
equation.  The lattice regularization of the quantum propagator is
defined in section \ref{Definition_of_Evolution_Kernels}.  There, we
also state the assumptions which are then used in section
\ref{Generalized_Schroedinger_Equation} for deriving the generalized
Schr\"odinger equation.  Both Hamilton-Jacobi and Sch\"odinger
equation are given in their integral form.  In the appendix we clarify
the relation with the local notation in \cite{Rovelli}.

\paragraph{Notation.}
$V$ is the spacetime domain over which the action and the path
integrals are defined.  $\Sigma$ is the boundary of $V$.  The letter $\phi$
denotes a real scalar field on $V$, while $\varphi$ stands for its
restriction to $\Sigma$, i.e.\ $\varphi = \phi|_{\Sigma}$.  Depending
on the context, $\phi$ can be a solution of the classical equations of
motion or an arbitrary field configuration.  The action associated to
$\phi$ is written as $S[\phi,V]$.  When boundary conditions
($\varphi$,$\Sigma$) determine a classical solution $\phi$ on $V$, we
denote the corresponding value of the action by $S[\varphi,V]$.  Thus,
the functional $S[\varphi,V]$ can be viewed as a Hamilton 
function (see sec.\ 3.3 of \cite{Rovelli}).  Vector components carry
greek indices $\mu, \nu, \ldots$ (e.g.\ $v = (v^\mu)$).  The dimension
of spacetime is $d$.  The symbol $\int_V \d^d x$ represents integrals
over $V$, while integrals over $\Sigma$ are indicated by
$\int_{\Sigma}\d\Sigma(x)$.  The letter $n$ denotes the outward
pointing and unit normal vector of $\Sigma$.  The normal derivative
is written as $\partial_n$, while $\naSig$ is the gradient along
$\Sigma$.  Accordingly, the full gradient $\nabla$ decomposes on
$\Sigma$ as 
\be
\label{decomposegradient}
\nabla|_\Sigma = n\,\partial_n + \naSig\,.
\ee
In section \ref{Definition_of_Evolution_Kernels}, we introduce a 
lattice with lattice spacing $a$ and regularize various continuum 
quantities. Their discrete analogues are designated by the index $a$: 
for example, $\varphi$, $V$ and $\Sigma$ become $\varphi_a$, $V_a$ 
and $\Sigma_a$.

\section{General Boundary Approach}
\label{General_Boundary_Approach}

What is the meaning of a state on a general surface which is not
necessarilly spacelike?  What does it mean to propagate fields along a
general spacetime domain?  Following \cite{Oeckl}, an intuitive answer
to these questions is provided by the path integral approach to QFT.
We illustrate this intuitive idea in this section, as a heuristic
motivation for the more rigorous definitions and developments in the
remainder of the article.  For simplicity, we refer here to a scalar
field theory, but similar considerations can be extended to various
path integral formulations of QFT, including sum over metrics or
spinfoam models in quantum gravity.

Consider Minkowskian scalar QFT in the Schr\"odinger picture.  Let 
$|\Psi_{\rm i}\k$ be an initial state at time $\ti$ and $|\Psi_{\rm 
f}\k$ a final state at time $\tf$. The transition amplitude between 
the two is 
\be 
A = \b\Psi_{\rm f}|\e^{-\i H(\tf-\ti)/\hbar}|\Psi_{\rm i}\k\,.
\label{amplitude}
\ee
Using the functional representation, the amplitude (\ref{amplitude})
can be expressed as a convolution 
\be
\label{transitionamplitude}
A = \int D\varphif\int D\varphii\, \Psi_{\rm 
f}^*[\varphif]\,W[\varphif,\tf;\varphii,\ti]\,\Psi_{\rm i}[\varphii]
\ee
with the propagator kernel
\be
W[\varphif,\tf;\varphii,\ti] := \b\varphif|\e^{-\i 
H(\tf-\ti)/\hbar}|\varphii\k\,.
\label{fieldproapagator}
\ee
This field propagator is a functional of the field: it should not be
confused with the Feynman propagator, which is a two-point function,
and propagates particles.  When rewritten as a path integral, this
kernel takes the form \be
\label{propagatortitf}
W[\varphif,\tf;\varphii,\ti] = \int\limits
_{\parbox{1.7cm}{\ssize $\phi(.,\ti)=\varphii\,,$ \\ 
$\phi(.,\tf)=\varphif$}}
 D\phi\;\e^{\i S[\phi,\ti,\tf]/\hbar}\,.
\ee
The action integral extends over the spacetime region $V_{\rm fi} := 
\bR^{d-1}\times [\ti,\tf]$ and the path integral sums over all field 
configurations $\phi$ on $V_{\rm fi}$ that coincide with the fields 
$\varphif$ and $\varphii$ on the boundary. The complete boundary 
consists of two parts: the hyperplane $\Sigma_{\rm i}$ at the initial 
time $\ti$, and the hyperplane $\Sigma_{\rm f}$ at the final time 
$\tf$. We call their union $\Sigma_{\rm fi}$ := $\Sigma_{\rm 
f}\cup\Sigma_{\rm i}$. If we view $\varphif$ and $\varphii$ as 
components of a single boundary field $\varphi_{\rm fi} := 
(\varphif,\varphii)$ on $\Sigma_{\rm fi}$, we can write the evolution 
kernel \eq{propagatortitf} more concisely as
$$
W[\varphi_{\rm fi},V_{\rm fi}] := \int_{\phi|_{\Sigma_{\rm 
fi}}=\varphi_{\rm fi}} D\phi\;\e^{\i S[\phi,V_{\rm fi}]/\hbar}\,.
$$
With this notation, it seems natural to introduce a propagator 
functional for more general spacetime regions $V$ (see 
\fig{fromVfitogeneralV}): we define it as 
\psfrag{ti}{$\ti$}\psfrag{tf}{$\tf$}
\psfrag{phii}{$\varphii$}\psfrag{phif}{$\varphif$}
\psfrag{Sigmaf}{$\Sigma_{\rm 
f}$}\psfrag{Sigmam}{$\Sigma_{\rm m}$}\psfrag{Sigmai}{$\Sigma_{\rm 
i}$}
\psfrag{V}{$V$}\psfrag{Sigma}{$\Sigma$}
\psfrag{phi}{$\varphi$}\pic{fromVfitogeneralV}{From 
$V_{\rm fi}$ to general $V$.}{4cm}{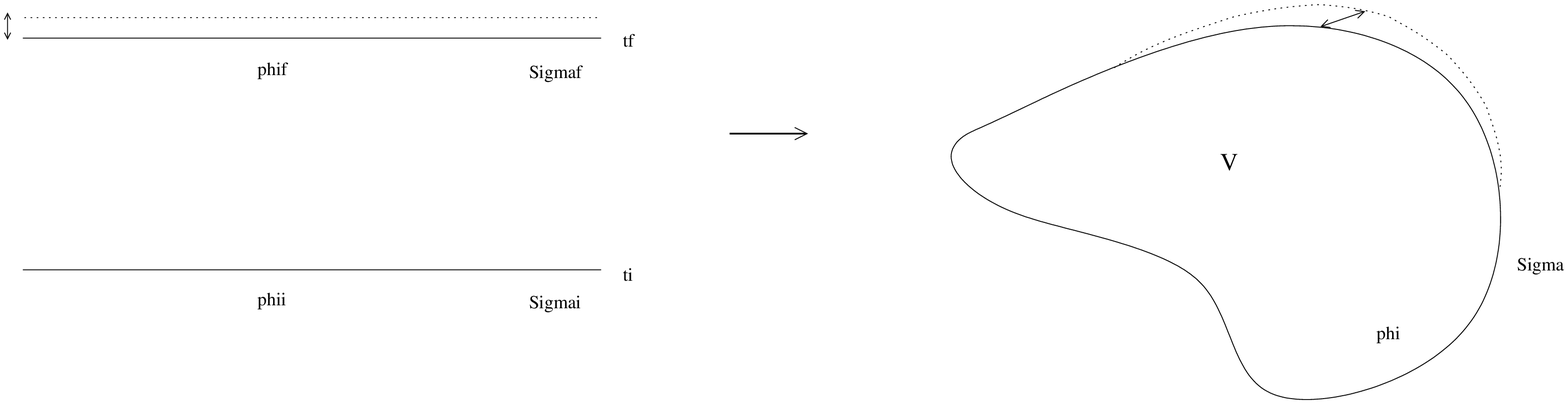}
\be
\label{propagatorV}
W[\varphi,V] := \int_{\phi|_\Sigma=\varphi} D\phi\;\e^{\i
S[\phi,V]/\hbar}\,.  \ee Here $\phi$ varies freely on the interior of
$V$ and is fixed to the value $\varphi$ on the boundary $\Sigma$.  Of
course, this is only a formal expression, and it is not clear that it
can be given mathematical meaning.  Let us suppose for the moment that
it {\it has} meaning and see what would follow from it.

\begin{figure}
\begin{center}
\psfrag{a}{$\varphi_c$}
\psfrag{b}{$\Sigma_c$}
\psfrag{c}{$\varphi_b$}
\psfrag{d}{$\Sigma_b$}
\psfrag{e}{$\varphi_a$}
\psfrag{f}{$\Sigma_a$}
\psfrag{g}{$V_{\rm ba}$}
\psfrag{h}{$V_{\rm cb}$}
\includegraphics[height=4cm]{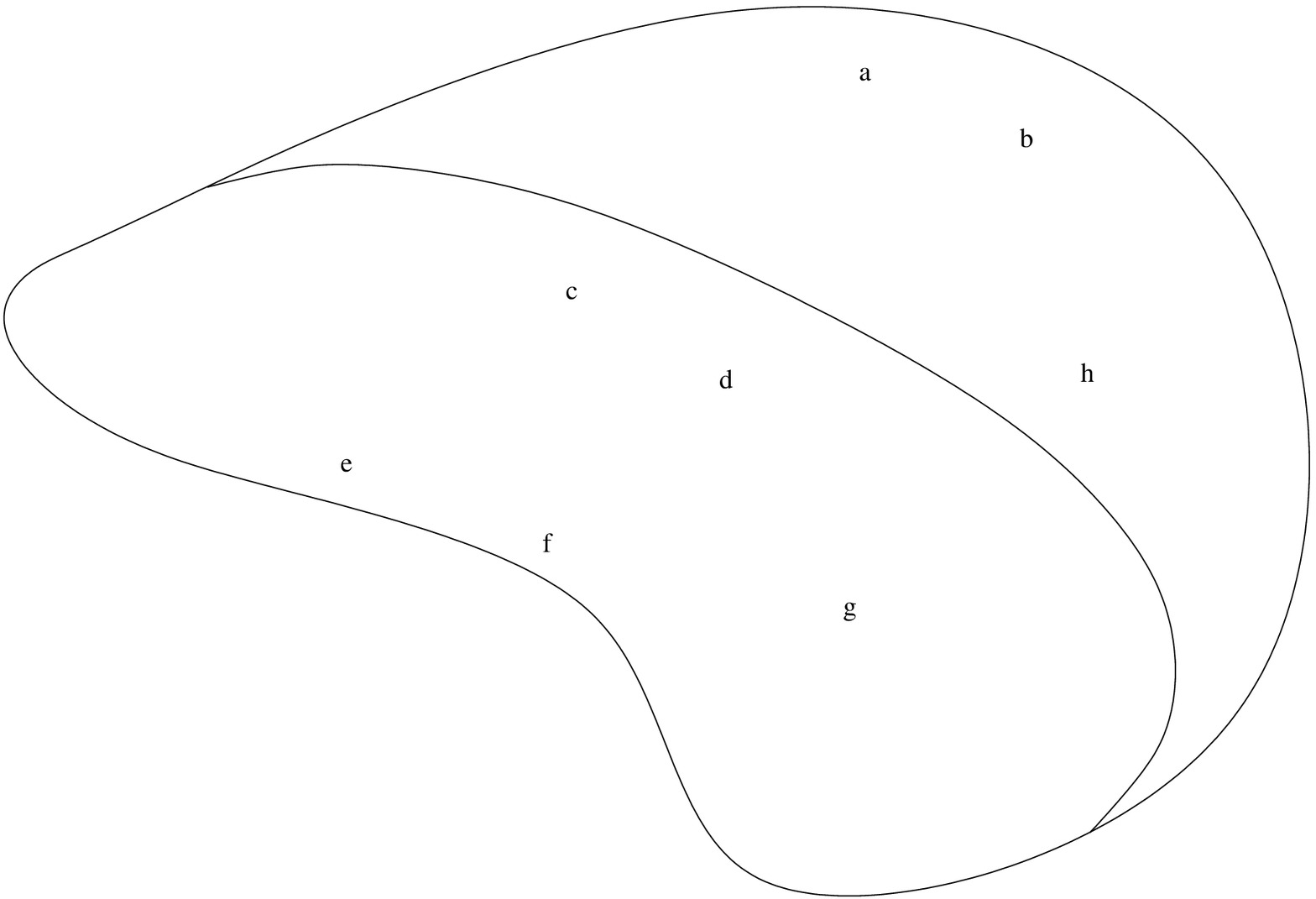}
\caption{\label{splittingofV} Splitting of $V$.}
\vspace{1cm}

\includegraphics[height=5cm]{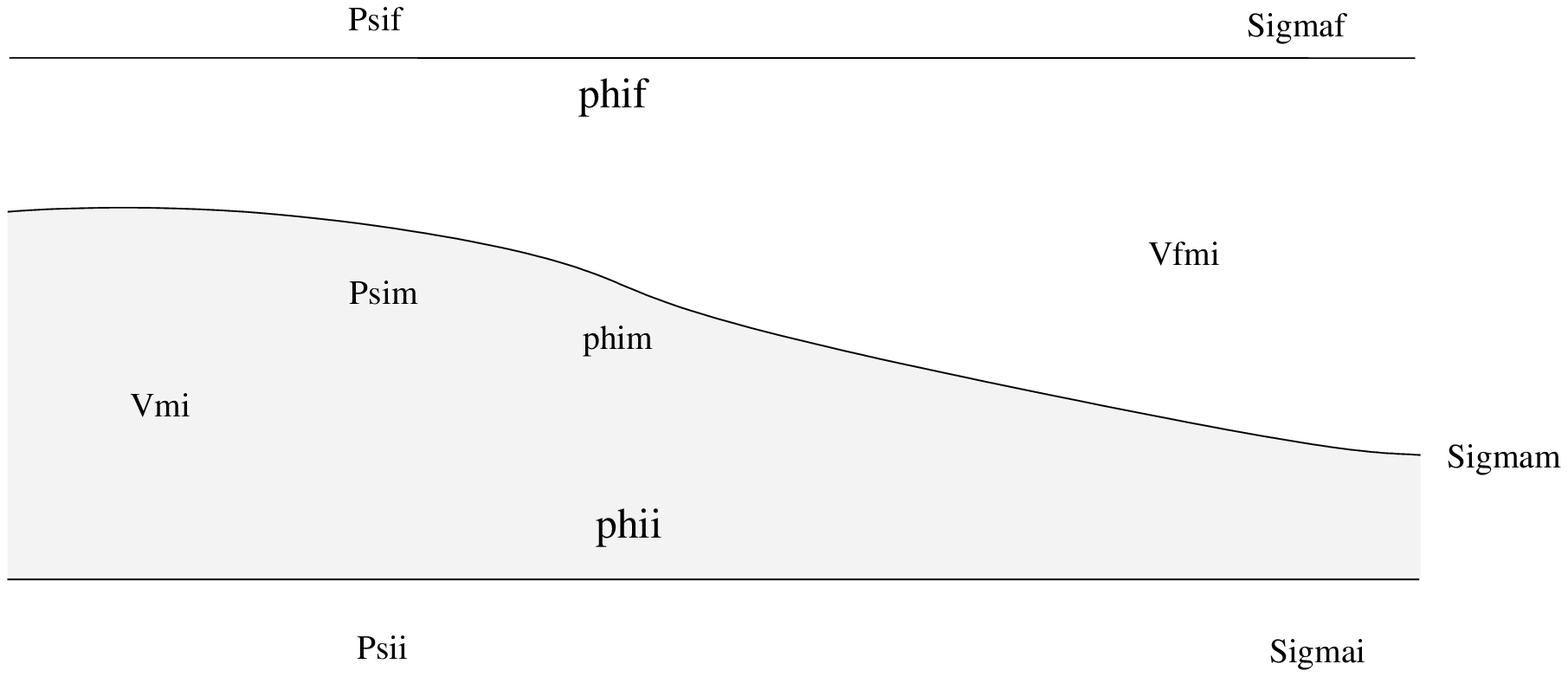}
\vspace{-5mm}

\caption{\label{splittingofVfi} Splitting of $V_{\rm fi}$.}
\end{center}
\end{figure}

Ordinary propagators satisfy convolution (or Markov) identities which 
result from the subdivision or joining of time intervals. If the 
functional $W$ behaves the way our naive picture tells us, the 
splitting and joining of volumes should translate into analogous 
convolution relations. For instance, if $V$ is divided as shown 
\fig{splittingofV}, the new regions $V_{\rm cb}$ and $V_{\rm ba}$ 
carry propagators
\bea
\label{propagatorVcb}
W[(\varphic,\varphib),V_{\rm cb}]&=&
\int\limits
_{\phi|_{\Sigma_{\rm cb}}=(\varphic,\varphib)}
D\phi\;\e^{\i S[\phi,V_{\rm cb}]/\hbar}\,, \\
\label{propagatorVba}
W[(\varphib,\varphia),V_{\rm ba}]&=&\int\limits_{\phi|_{\Sigma_{\rm 
ba}}=(\varphib,\varphia)} D\phi\;\e^{\i S[\phi,V_c]/\hbar}\,.
\eea
When integrating the product of \eq{propagatorVcb} and 
\eq{propagatorVba} over the field $\varphi_{\rm b}$ 
along the common boundary, one 
recovers the original propagator: \psfrag{Psif}{$\Psi_{\rm f}$}
\psfrag{Psii}{$\Psi_{\rm i}$}
\psfrag{Psim}{$\Psi_{\rm m}$}
\psfrag{phim}{$\varphim$}
\psfrag{Vm}{$V_{\rm m}$}
\psfrag{Vfmi}{$V_{\rm fm}$}
\psfrag{Vmi}{$V_{\rm mi}$}
\be
\label{generalizedconvolutionrelation}
W[(\varphic,\varphia),V] = \int 
D\varphib\,W[(\varphic,\varphib),V_{\rm 
cb}]\,W[(\varphib,\varphia),V_{\rm ba}]\,.
\ee
Similarly, the infinite strip $V_{\rm fi}$ between $\tf$ and $\ti$ 
could be cut by a ``middle'' surface $\Sigma_{\rm m}$  as in 
\fig{splittingofVfi}, giving the new volumes $V_{\rm fm}$ and $V_{\rm 
mi}$. Its kernel decomposes as 
$$W[(\varphif,\varphii),V_{\rm fi}] = \int 
D\varphim\,W[(\varphif,\varphim),V_{\rm 
fm}]\,W[(\varphim,\varphii),V_{\rm mi}]\,.$$
Thus, the evolution of $|\Psi_{\rm i}\k$ to the final time is divided 
into two steps: using the propagator on $V_{\rm mi}$, we evolve up to 
the surface $\Sigma_{\rm m}$ and obtain the intermediate state
$$\Psi_{\rm m}[\varphim] := 
\int\d\varphii\,W[(\varphim,\varphii),V_{\rm mi}]\,\Psi_{\rm 
i}[\varphii]\,.$$
The kernel $W[.,V_{\rm fm}]$ covers the remaining evolution and gives 
the original amplitude \eq{amplitude} when convoluted with 
$\Psi_{\rm f}^*$ and $\Psi_{\rm m}$
\be
\label{amplitudePsim}
A = \int D\varphif\int D\varphim\, \Psi_{\rm 
f}^*[\varphif]\,W[(\varphif,\varphim),V_{\rm fm}]\,\Psi_{\rm 
m}[\varphim]\,.
\ee
Since the same amplitude can either be calculated from $\Psi_{\rm f}$ 
and $\Psi_{\rm i}$ or from $\Psi_{\rm f}$ and $\Psi_{\rm m}$, the 
wave functional $\Psi_{\rm m}$ encodes all physical information about 
the intial state. On account of this property, we say that {\it 
$\Psi_{\rm m}$ is the state functional which results from evolving 
$\Psi_{\rm i}$ by the volume $V_{\rm mi}$}. As for ordinary state 
functionals, one can think of $\Psi_{\rm m}$ as being an element 
$|\Psi_{\rm m}\k$ in a Hilbert space, which we call $\H_{\Sigma_{\rm 
m}}$. The latter consists of functionals of fields over $\Sigma_{\rm 
m}$ and has the inner product
$$\b\Psi_2|\Psi_1\k := \int 
D\varphim\,\Psi^*_2[\varphim]\Psi_1[\varphim]\,,\qquad 
|\Psi_1\k\,,\,|\Psi_2\k\in\H_{\Sigma_{\rm m}}\,.$$
It is important to note that the evolution map from $\H_{\Sigma_{\rm 
i}}$ to $\H_{\Sigma_{\rm m}}$ need not be unitary. The results of 
Torre and Varadarajan show, in fact, that in flat spacetime state 
evolution between curved Cauchy surfaces cannot be implemented 
unitarily \cite{Torre}. Nevertheless, a probability interpretation 
{\it is} viable for states in $H_{\Sigma_{\rm m}}$, as the meaning of 
amplitudes such as \eq{amplitudePsim} can be traced back to that of 
the standard amplitude \eq{transitionamplitude}.

Consider now a more unconventional example. Cut out a bounded and
simply connected set $V_{\rm m}$ from $V_{\rm fi}$ and denote the
remaining volume by $V_{\rm fmi}$ (\fig{hole}).  \psfrag{Vfmi}{$V_{\rm
fmi}$}\pic{hole}{Evolution to a closed surface $\Sigma_{\rm
m}$.}{4cm}{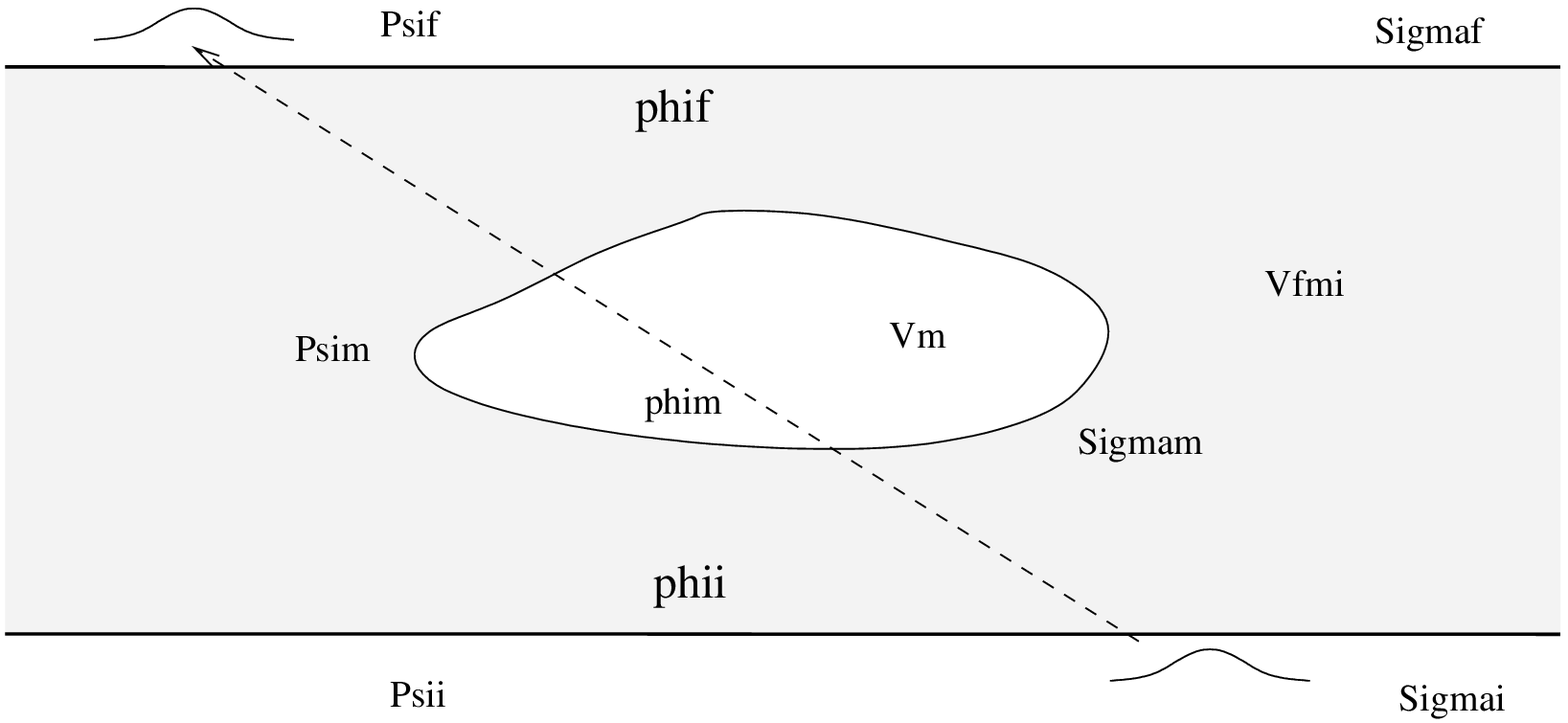} This time we define the state functional
$\Psi_{\rm m}$ by ``evolving'' both $\Psi_{\rm f}$ and $\Psi_{\rm m}$
to the middle boundary $\Sigma_{\rm m}$, i.e. $$\Psi_{\rm m}[\varphim]
:= \int D\varphif\int D\varphii\, \Psi_{\rm
f}^*[\varphif]\,W[(\varphif,\varphim,\varphii),V_{\rm fmi}]\,\Psi_{\rm
i}[\varphii]\,.
$$
Clearly, the amplitude \eq{transitionamplitude} is now equal to
\be
\label{holeamplitude}
A = \int D\varphim\,W[\varphim,V_{\rm m}]\,\Psi_{\rm m}[\varphim]\,. 
\ee 
Therefore, the functional $\Psi_{\rm m}$ contains the entire
information needed to compute the transition amplitude between
$\Psi_{\rm i}$ and $\Psi_{\rm f}$.

To make this more concrete and more intuitive, suppose that the scalar
field theory is free and that $\Psi_{\rm i}$ and $\Psi_{\rm f}$ are
the initial and final one-particle states of a single, localized
particle whose (smeared out) worldline passes through $V_{\rm m}$.  In
both functionals, the presence of the particle appears as a local
deviation from the vacuum, in the functional dependence.  Likewise, it
is natural to presume that the functional form of $\Psi_{\rm m}$
reflects where the worldline of the particle enters and exits the
volume $V_{\rm m}$.

How can we interpret the ``state" $\Psi_{\rm m}$ and the
associated amplitude \eq{holeamplitude}?  To answer this, let us get back
to equation \eq{amplitude}.  Notice that the amplitude $A$ depends on
the \emph{couple} of states $(|\Psi_{\rm i}\k,|\Psi_{\rm f}\k)$.  This
couple represents a possible outcome of a measurement at time $\tf$ as
well as a state preparation at time $\ti$.  A state preparation is
itself a quantum measurement, therefore we can say that this couple
represents a possible outcome of an ensemble of quantum measurements
performed at times $\ti$ and $\tf$.  We may introduce a name to denote
such a couple.  We call it a \emph{process}, since the two states
$(|\Psi_{\rm i}\k,|\Psi_{\rm f}\k)$, taken together, represent the
ensemble of data (initial and final) that we can gather about a
physical process.  A probability amplitude is associated to the entire
process $(|\Psi_{\rm i}\k,|\Psi_{\rm f}\k)$.  Now, it is clear that
the functional $\Psi_{\rm m}$ represents a generalization of this idea
of a process.  It is tempting to presume that $\Psi_{\rm m}$ can be
interpreted as representing a possible outcome of quantum measurements
that can be made on $\Sigma_{\rm m}$.  In the example of the particle
above, for instance, it will represent the detection of the incoming
and outgoing particle.

The idea is that given an arbitrary closed surface, the possible
results of the ensemble of measurements that we can make on it
determines a space of generalized ``states'' which can be associated to
the surface.  Each such state represents a process whose
probabilistic amplitude is provided by expression \eq{holeamplitude}.
The conventional formalism is recovered when the surface is formed by two
parallell spacelike planes.  For more details on the physical
interpretation of general boundary states, see sec.\ 5.3 of
\cite{Rovelli}.

\subsection{Operator Formalism}
\label{Operator_Formalism}

If path integrals can be defined for general boundaries, how would a
corresponding operator formalism look like?  In particular, is there
an operator that governs the dynamics, as the Hamiltonian does for
rigid time translations?  Recall that the Hamiltonian can be recovered
from the path integral by considering an infinitesimal shift of the
final time.  For example, if we displace by a time interval $\Delta t$
the final surface $\Sigma_{\rm f}$ in \eq{propagatortitf}, keeping the
same boundary field $\varphif$, the new propagator results from the
convolution 
\be
W[\varphif,\tf+\Delta t;\varphii,\ti] = \int
D\varphi\,W[\varphif,\tf+\Delta
t;\varphi,\tf]\,W[\varphi,\tf;\varphii,\ti]\,.
\ee
For infinitesimal $\Delta t$, this gives the Schr\"odinger equation, 
which expresses the variation of $W$ in terms of the Hamiltonian  
operator
\be
\label{Schroedingerequation}
\left(\i\hbar\pdiff{}{\tf} - 
H[\varphif,-\i\hbar\fdiff{}{\varphif}]\right)W[\varphif,\tf;\varphii,\ti] 
= 0\,, 
\ee
where
\be
H[\varphif,\pif] \ =\  \int_{\Sigma_{\rm 
f}}\d\Sigma\  \frac{1}{2} \left(\pif^2 + (\naSig\varphif)^2 
+ m^2\varphif^2\right)\,. \nonumber
\ee
Similarly, if $\varphif$ is displaced in a tangential direction 
$e_\parallel$ along $\Sigma_{\rm f}$, the variation of $W$ is 
generated by the momentum operator
$$e_\parallel\cdot P[\varphif,-\i\hbar\fdiff{}{\varphif}]\,,$$
where
$$P[\varphif,\pif] = -\int_{\Sigma_{\rm 
f}}\d\Sigma(x)\,\naSig\varphif(x)\,\pif\,.$$

In the case of a general volume $V$, it is natural to expect that
deformations of the boundary surface $\Sigma$ lead to an analogous
functional differential equation for the propagator.  However, for a general
shape of $V$ there is no notion of preferred rigid displacement of
the boundary.  We must consider arbitrary deformations of the
boundary surface, and we expect that the associated change in $W$ is governed by a generalized Schr\"odinger equation (see \fig{fromVfitogeneralV}).  In the same way that $H$ and $P$ generate temporal and spatial shifts, the operators in such a Schr\"odinger equation could be seen as the generators for general boundary deformations of $W$. In a difformopshim invariant QFT, the analogous $W$-functional would be independent of $\Sigma$, and the generalized Schr\"odinger equation reduces to the Wheeler-DeWitt equation.

In the present paper, we consider only Euclidean field theory, so we
seek to define the Euclidean form 
\be
\label{EuclideanpropagatorV}
W[\varphi,V] := \int_{\phi|_\Sigma=\varphi} D\phi\;\e^{- 
S[\phi,V]/\hbar}\,.
\ee
of the propagator \eq{propagatorV}, and generalize the Euclidean 
version of the Schr\"odinger equation \eq{Schroedingerequation}.

Before dealing with path integrals and deformations of their
boundaries, however, we discuss the analagous problem in classical
field theory.  The classical counterpart of the Schr\"odinger equation
is the Hamilton-Jacobi equation.  The Hamilton function
$S[\varphif,\tf,\varphii,\ti]$ is a function of the same arguments as
the field propagator \eq{fieldproapagator}.  It is defined as the
value of the action of the classical field configuration which solves
the equations of motion and has the given boundary values.  It
satisfies the Hamilton-Jacobi equation \be
\pdiff{}{\tf}S[\varphif,\tf;\varphii,\ti] +
H[\varphif,\fdiff{S}{\varphif}] = 0.  
\ee
For more general regions $V$, the Hamilton function becomes 
a functional of $V$ and the boundary field $\varphi$ specified on 
$\Sigma$. In the next section we show that this functional satisfies 
a generalized Hamilton-Jacobi equation which governs its dependence on
arbitrary variations of $V$.

\section{Generalized Hamilton-Jacobi Equation}
\label{Generalized_Hamilton_Jacobi_Equation}

Let $V$ be an open and simply connected subset of Euclidean 
$d$-dimensional space $\bR^d$. We consider the Euclidean action 
\be
S[\phi,V]=\int_V d^d x 
\left[\frac{1}{2}(\nabla\phi)^2+\frac{1}{2}m^2\phi^2 + 
U(\phi)\right]\,,
\ee
where $U$ is some polynomial potential in $\phi$. In the classical 
case, unlike in the quantum case, an interaction term can be added 
without complicating the derivation that follows. The equations of 
motion are
\be
\label{eqnsofmotion}
\Box\phi - m^2\phi - \pdiff{U}{\phi} = 0\,.
\ee
The Hamilton function $S[\varphi,V]$ is defined by
$S[\varphi,V]=S[\phi,V]$, where $\phi$ solves \eq{eqnsofmotion} and
$\phi|_{\Sigma}=\varphi$.  It is defined for all values
$(\varphi,\Sigma)$ where this solution exists and is multivalued if
this solution is not unique.

We now study the change in $S[\varphi,V]$ under a local 
variation of $V$. To make this precise, consider a vector field $N = 
(N^\mu)$ over $\bR^d$. $N$ induces a flow on $\bR^d$ which we denote 
by $\sigma: \bR\times\bR^d\to \bR^d$. Define the transformed volume 
as $V^s := \sigma(s,V)$. Likewise, $\Sigma^s := \sigma(s,\Sigma)$. 

To define the change in $S[\varphi,V]$ under a variation of $V$, we
need to specify what value the boundary field should take on the new
boundary $\Sigma^s$.  We choose it to be the pull-forward by $\sigma_s
\equiv \sigma(s,.)$, i.e.\ $\varphi^s := \varphi\circ\sigma_s^{-1}
\equiv \sigma_{s*}\varphi$.  Let us assume that the point
$(\varphi,\Sigma)$ is regular in the space of boundary conditions, in
the sense that slightly deformed boundary conditions
$(\varphi^s,\Sigma^s)$ give a new unique solution $\phi^s$ on $V^s$,
close to the previous one.  In this case, the number
$S[\varphi^s,V^s]$ is well-defined and we can write down
the differential quotient 
\be
L_NS[\varphi,V] := \lim_{s\to 0}\frac{1}{s}\left(S[\varphi^s,V^s] - 
S[\varphi,V]\right)\,,
\label{L}
\ee
with the vector field $N$ as a parameter. As we show below, this limit exists and the map $L_N$ is a functional differential operator. The local form of this differential operator is given in the appendix. 

We decompose the restriction of $N$ to $\Sigma$ into its components
normal and tangential to $\Sigma$,
$$N_{|\Sigma} = N_\perp n + N_\parallel\,,$$
where the scalar field $N_\perp$ is defined as
$$N_\perp := n_\mu N^\mu\,.$$  
Observe that under a small variation $\delta\varphi$ of the boundary
field, we have
\bean 
\lefteqn{\delta S[\varphi,V] = S[\delta\varphi,V] =
S[\delta\phi,V]} \\
&&= \int_{V} d^d x\left[\nabla^\mu\delta\phi\,\nabla_{\!\mu}\phi +
m^2\phi\,\delta\phi + \pdiff{U}{\phi}\delta\phi\right] \\
&&= \int_{V} d^d
x\bigg[\nabla_{\!\mu}(\nabla_{\!\mu}\phi\,\delta\phi) +
\delta\phi\underbrace{\left(-\Box\phi + m^2\phi + 
\pdiff{U}{\phi}\right)}_{=
0}\bigg] \\
 &&= \int_\Sigma\d\Sigma\,\partial_n\phi\,\delta\phi
= \int_\Sigma\d\Sigma\,\,\partial_n\phi\,\delta\varphi\,.
\eean
Therefore, we have
\be
\label{Soverphi}
\fdiff{S}{\varphi(x)}[\varphi,V] = \partial_n\phi(x)\,.
\ee

\subsection{Direct Derivation}
\label{Direct_Derivation}
Suppose for the moment that $V$ is only extended by the deformation 
(i.e.\ $V\subset V^s$ for every $s$). Then, the most direct 
derivation of the Hamilton-Jacobi equation can be obtained by 
considering the restriction $\varphi^s_\Sigma$ of $\phi^s$ to 
$\Sigma$: that is, the value of the classical solution on $\Sigma$ 
when the boundary condition $\varphi^s$ is specified on $\Sigma^s$ 
(see \fig{phisSigma}). 
\psfrag{phisSigma}{$\varphi^s_\Sigma$}\psfrag{Vs}{$V^s$}
\psfrag{Sigmas}{$\Sigma^s$}\psfrag{phis}{$\varphi^s$}
\psfrag{phissol}{$\phi^s$}

\pic{phisSigma}{Definition of 
$\varphi^s_\Sigma$.}{4cm}{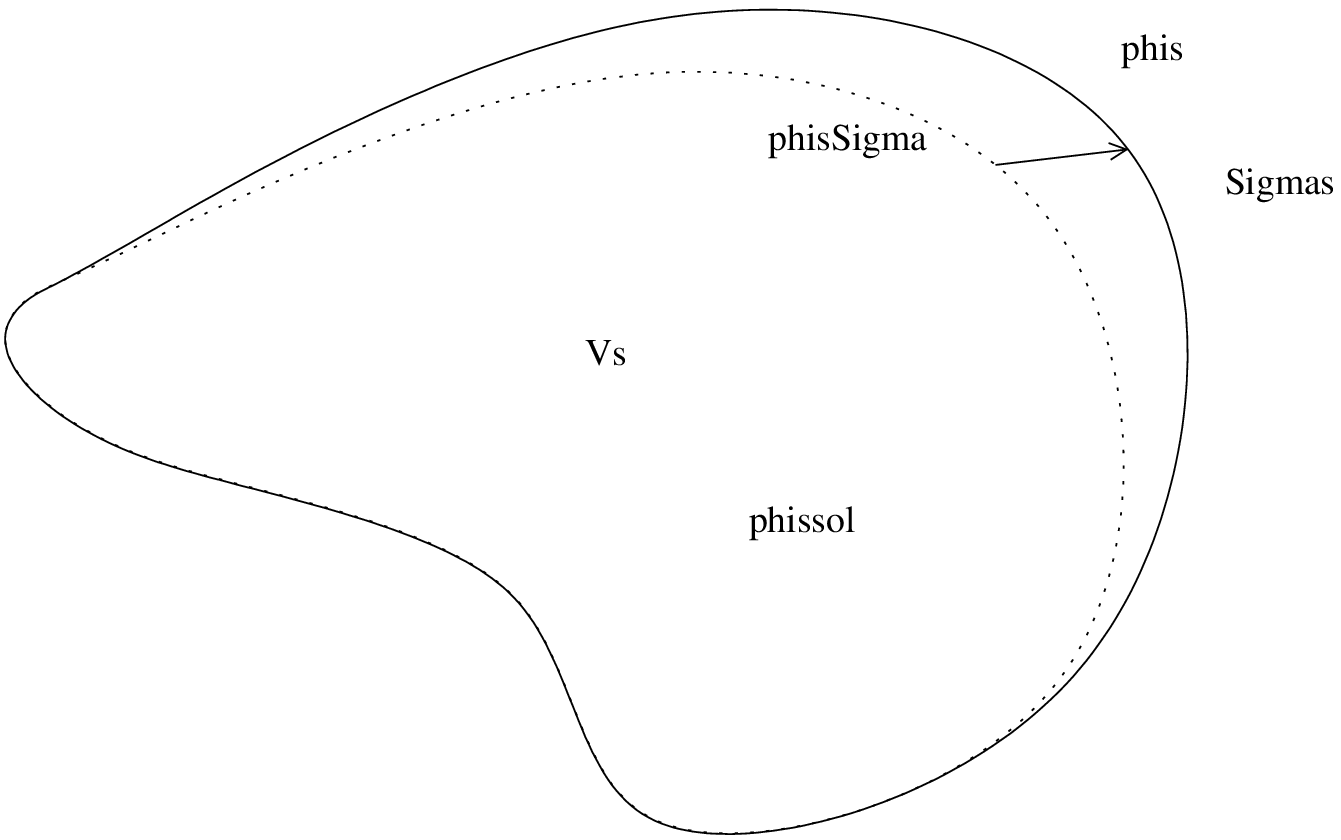}  
Note that $\varphi^0_\Sigma = \varphi^0 = \varphi$. By inserting 
$S[\varphi^s_\Sigma,V]$ into the difference, i.e.
$$S[\varphi^s,V^s] - S[\varphi,V]
= S[\varphi^s,V^s] - S[\varphi^s_\Sigma,V]
+ S[\varphi^s_\Sigma,V] - S[\varphi,V]\,,$$
the differential quotient becomes a sum of two limits:
$$\lim_{s\to 0} \frac{S[\varphi^s,V^s] - S[\varphi,V]}{s}
= \lim_{s\to 0}\frac{S[\varphi^s,V^s] - S[\varphi^s_\Sigma,V]}{s} 
+ \lim_{s\to 0}\frac{S[\varphi^s_\Sigma,V] - S[\varphi,V]}{s}$$
As $\varphi^s$ and $\varphi^s_\Sigma$ are part of the same solution, 
the first limit is easily seen to be
\bean
\lim_{s\to 0} \frac{S[\varphi^s,V^s] - S[\varphi^s_\Sigma,V]}{s}
&=&
\int_{\Sigma}\d\Sigma\;N_\perp\!\left(\frac{1}{2}
\nabla^\mu\phi\nabla_{\!\mu}\phi 
+
\frac{1}{2}m^2\phi^2 + U(\phi)\right) \\
 &=&
\int_{\Sigma}\d\Sigma\;N_\perp\!\left(\frac{1}{2}
\left(\fdiff{S}{\varphi}\right)^2 
+ \frac{1}{2}(\naSig\phi)^2 + \frac{1}{2}m^2\phi^2 + U(\phi)\right)\,.
 \eean
 In the last line, we used the decomposition \eq{decomposegradient} 
and equation \eq{Soverphi}. The second differential quotient gives
\bean
\lim_{s\to 0}\frac{S[\varphi^s_\Sigma,V] - S[\varphi,V]}{s}
&=& 
\int_{\Sigma}\d\Sigma\;\fdiff{S}{\varphi}\,
\diff{}{s}\varphi^s_\Sigma|_{s=0} 
\nonumber \\
&=& \int_{\Sigma}\d\Sigma\;\fdiff{S}{\varphi}\left(- 
N_\perp\partial_n\phi -
N_\parallel\cdot\naSig\varphi\right)\,.
 \eean
Altogether one has
\be
\label{beforeintroducingHNPN}
L_NS[\varphi,V] =
\int_{\Sigma}\d\Sigma\left\{N_\perp\left[-\frac{1}{2}
\left(\fdiff{S}{\varphi}\right)^2 
+ \frac{1}{2}(\naSig\phi)^2 + \frac{1}{2}m^2\phi^2 + U(\phi)\right] - 
N_\parallel\cdot\naSig\varphi\fdiff{S}{\varphi}\right\}\,.
\ee
We started from the assumption that $V\subset V^s$ for all $s$, but 
it is easy to see that the previous argument can be adapted to the 
general case where the volume $V$ is partly extended and partly 
decreased. 

If we introduce the
quantities
\bean
H_N[\varphi,\pi,V] &:=& 
\int_\Sigma\d\Sigma\;N_\perp\!\left(-\frac{1}{2}\pi^2 +
\frac{1}{2}(\naSig\varphi)^2 + \frac{1}{2}m^2\varphi^2 + 
U(\varphi)\right)\,,
\\
P_N[\varphi,\pi,V] &:=&
-\int_\Sigma\d\Sigma\,N_\parallel\cdot\naSig\varphi\,\pi\,,
\eean
equation
\eq{beforeintroducingHNPN} takes the form
\be
\label{generalizedHJ}
L_NS[\varphi,V] = H_N[\varphi,\fdiff{S}{\varphi},V] + 
P_N[\varphi,\fdiff{S}{\varphi},V]\,.
\ee
When $V$ is a strip of spacetime between times $\ti$ and $\tf$, and 
$N_\perp|_{\tf} = 1$, $N_\perp|_{\ti} = 0$, $N_\parallel = 0$, 
equation \eq{generalizedHJ} reduces to the usual Hamilton-Jacobi 
equation
$$\pdiff{}{\tf}S[\varphif,\tf;\varphii,\ti] = \int_{\Sigma_{\rm 
f}}\d\Sigma\,\left(-\frac{1}{2}\left(\fdiff{S}{\varphif}\right)^2 + 
\frac{1}{2}(\naSig\varphif)^2 + \frac{1}{2}m^2\varphif^2 + 
U(\varphif)\right)$$
of Euclidean field theory. Hence \eq{generalizedHJ} can be seen as a 
geometric generalization of the Hamilton-Jacobi equation. 

\subsection{Alternative Derivation}
\label{Alternate_Derivation}
Let us describe another way of evaluating the ``deformation
derivative'' $L_{N}S[\varphi,V]$.  The spacetime metric tensor $g$
enters in the definition of the action and therefore in the definition
of $S[\varphi,V]$.  Let us write this dependence explicitely as
$S[\varphi,g,V]$.
A diffeomorphism that acts on $\phi$, the boundary $\Sigma$ 
{\it and} the metric $g$, leaves the action invariant. Therefore
$$
S[\varphi^s,g^s,V^s] = S[\varphi,g,V]\,.
$$
Equivalently,
$$
S[\varphi^s ,g,V^s] = S[\varphi,g^{-s},V] .
$$
Plugging this into the definition of the operator \eq{L} gives \be
\label{onlyvariationofg}
L_{N}S[\varphi,g,V] = \lim_{s\to 0} \frac{1}{s}\left(S[\varphi,g^{-s},V] - 
S[\varphi,g,V]\right)\,,
\ee
which is a variation of the action w.r.t.\ the metric {\it only}. Now 
we can use the definition of the energy-momentum tensor to obtain
\bea
L_{N}S[\varphi,g,V]  &=& \frac{1}{2}\int_V d^d
x\,T^{\mu\nu}\diff{}{s}g^{-s}_{\mu\nu}|_{s=0} = \frac{1}{2}\int_V d^d
x\,T^{\mu\nu}(-\nabla_{\!\mu} N_\nu-\nabla_{\!\nu} N_\mu) \nonumber \\
&=& -\int_V d^d
x\,T^{\mu\nu}\nabla_{\!\mu} N_\nu = -\int_{\Sigma}\d\Sigma\,n_{\mu} 
T^{\mu\nu}N_\nu +
\int_V\d^d x\underbrace{\nabla_{\mu}T^{\mu\nu}}_{=0} N_\nu \nonumber 
\\
&=&
-\int_{\Sigma}\d^d x\,n_{\mu} T^{\mu\nu}N_\nu 
 \label{intermediate}
\eea 
In the last two steps we used Stoke's theorem and the equations
of motion respectively.
On the other hand, we know that
$$
T^{\mu\nu} =
-g^{\mu\nu}\L + \nabla^\mu\phi\nabla^\nu\phi
$$
and
\bea
n_{\mu}N_{\nu} T^{\mu\nu} &=& 
-N_\perp\left[\frac{1}{2}(\partial_n\phi)^2 +
\frac{1}{2}(\naSig\phi)^2 + \frac{1}{2}m^2\phi^2 + U(\phi)\right] 
+ \partial_n\phi(N_\perp\pa_n\phi + N_\parallel\cdot\naSig\phi)
\nonumber \\
&=& -N_\perp\left[-\frac{1}{2}(\partial_n\phi)^2 +
\frac{1}{2}(\naSig\phi)^2 + \frac{1}{2}m^2\phi^2 + U(\phi)\right]
+ N_\parallel\cdot\naSig\phi\,\partial_n\phi
\label{vectorsEMT}
\eea
Inserting \eq{vectorsEMT} in \eq{intermediate} and
using \eq{Soverphi}, we arrive again at the generalized 
Hamilton-Jacobi
equation
$$
L_N S[\varphi,V] =
\int_{\Sigma}\d\Sigma\left\{N_\perp\left[-\frac{1}{2}
\left(\fdiff{S}{\varphi}\right)^2 
+ \frac{1}{2}(\naSig\phi)^2 + \frac{1}{2}m^2\phi^2 + U(\phi)\right] - 
N_\parallel\cdot\naSig\varphi\fdiff{S}{\varphi}\right\}\,.
$$

\section{Definition of the Evolution Kernel}
\label{Definition_of_Evolution_Kernels}

In this section, we define a Euclidean free field propagator for
arbitrary spacetime domains $V$.  Limits of lattice path integrals are
used to give a precise meaning to the expression
\eq{EuclideanpropagatorV}.  We begin by considering the case $V =
V_{\rm fi}$ and derive the lattice path integral from the operator
formalism.  Then, we propose a way to extend this expression to more
general volumes $V$.

\subsection{From Operators to Path Integrals}
\label{From_Operators_to_Path_Integrals}

The transition from operator formalism to path integral is a standard 
procedure. We repeat it here, since treatments of lattice field 
theory usually omit normalization factors. There, constant factors 
drop out when dividing by the partition function $Z$. In our case, 
their precise form will be crucial for the definition of the 
propagator.  

In the Schr\"odinger picture, the space of states $\H$ is associated 
to the manifold $\bR^{d-1}$: we regularize it by a finite lattice 
$$S_a := \{\v{x}\in a\bZ^{d-1}\;|\;-Ma\le |x_i|\le 
Ma\,,\;i=1,\ldots,d-1\} $$ 
with lattice constant $a>0$ and edge length $2aM$, $M\in\bN$. $e_i$ 
is the unit vector in the $i$th direction. For a scalar function $f$ 
on $S_a$, the forward derivative is
$$\nabla_i f(\v{x}) := \frac{\phi(\v{x}+a e_i)-\phi(\v{x})}{a}\,,$$
and we set $\phi(\v{x}+a e_i) := \phi(\v{x}-aM e_i)$ when $x_i = aM$.
Let $\{\hat{\phi}(\v{x})\}$, $\{\hat{\pi}(\v{x})\}$ be canonical 
operators with eigenstates $\{|\phi\k\}$, $\{|\pi\k\}$ such that
\be
\label{eigenstates}
\hat{\phi}(\v{x})\,|\,\phi\k = 
\phi(\v{x})|\phi\k,\,\qquad\hat{\pi}(\v{x})|\pi\k = 
\pi(\v{x})|\pi\k\,,\qquad \v{x}\in S_a\,,
\ee
and
\be
\label{commutationrelations}
[\hat{\phi}(\v{x}),\hat{\pi}(\v{y})] = 
\frac{\i\hbar}{a^{d-1}}\delta(\v{x}-\v{y})\,,\qquad \v{x},\v{y}\in 
S_a\,.
\ee
The eigenstates are normalized as
\be
\label{normalization}
\b\phi,\phi'\k = \prod_{\v{x}\in 
S_a}\delta(\phi(\v{x})-\phi'(\v{x}))\,,\qquad \b\pi,\pi'\k = 
\prod_{\v{x}\in S_a}\delta(\pi(\v{x})-\pi'(\v{x}))\,,
\ee
and give rise to completeness relations
\be
\label{completeness}
\left(\prod_{\v{x}\in 
S_a}\int_{-\infty}^\infty\d\phi(\v{x})\right)|\phi\k\b\phi| = 
\mathbbm{1}\,,\qquad\left(\prod_{\v{x}\in 
S_a}\int_{-\infty}^\infty\d\pi(\v{x})\right)|\pi\k\b\pi| = 
\mathbbm{1}\,.
\ee
From \eq{eigenstates}, \eq{commutationrelations} and 
\eq{normalization}, it follows that 
\be
\label{momentumoperator}
\hat{\pi}(\v{x}) = -\frac{\i\hbar}{a^{d-1}}\pdiff{}{\phi(\v{x})}
\ee
and
\be
\label{productphipi}
\b\phi,\pi\k = \left(\prod_{\v{x}\in 
S_a}\sqrt{\frac{a^{d-1}}{2\pi\hbar}}\right)
\exp\left(\frac{\i}{\hbar}\sum_{\v{x}\in 
S_a}a^{d-1}\phi(\v{x})\pi(\v{x})\right)
\ee
The Hamiltonian operator is
\bean
H[\hat{\phi},\hat{\pi}] &:=& \sum_{\v{x}\in 
S_a}a^{d-1}\left[\frac{1}{2}\hat{\pi}^2(\v{x}) + 
\frac{1}{2}(\v{\nabla}\hat{\phi})^2(\v{x}) + 
\frac{1}{2}m^2\hat{\phi}^2(\v{x})\right] \\
&\equiv& T[\hat{\pi}] + V[\hat{\phi}]\,.
\eean

We rewrite the Euclidean propagator 
$$ 
\b\varphif|\e^{-H(\tf-\ti)/\hbar}|\varphii\k\,,\qquad \tf-\ti = 
na\,,$$
by inserting repeatedly the completeness relations \eq{completeness}:
\bean
\lefteqn{\b\varphif|\e^{-na H/\hbar}|\varphii\k
= \left(\prod_{k=1}^{n-1}\prod_{\v{x}}\int\d\phi_k(\v{x})\right)
\left(\prod_{k=0}^{n-1}\prod_{\v{x}}\int\d\pi_k(\v{x})\right)\,\times} 
\\
&&\times\,\b\varphif|\pi_{n-1}\k\b\pi_{n-1}|\e^{-a 
H/\hbar}|\phi_{n-1}\k
\b\phi_{n-1}|\pi_{n-2}\k\b\pi_{n-2}|\e^{-a H/\hbar}|\phi_{n-2}\k
\cdots 
\b\phi_1|\pi_0\k\b\pi_0|\e^{-a H/\hbar}|\varphii\k
\eean
After making the replacement
$$\e^{-a H/\hbar} =  \e^{-a T/\hbar}\e^{-a V/\hbar} + O(a^2)\:\to\: 
\e^{-a T/\hbar}\e^{-a V/\hbar}$$
and using \eq{productphipi}, we obtain
\bean
\lefteqn{\left(\prod_{k=1}^{n-1}
\prod_{\v{x}}\int\d\phi_k(\v{x})\right)
\left(\prod_{k=0}^{n-1}\prod_{\v{x}}\int\d\pi_k(\v{x})\right)
\b\varphif|\pi_{n-1}\k\b\pi_{n-1}|\phi_{n-1}\k
\b\phi_{n-1}|\pi_{n-2}\k\b\pi_{n-2}|\phi_{n-2}\k
\cdots} \\
&& \hspace{5.8cm}\cdots\b\phi_1|\pi_0\k\b\pi_0|\varphii\k
\exp\left(\frac{1}{\hbar}\sum_{k=0}^{n-1}
aH[\phi_k,\pi_k]\right)_{\Big|\parbox[b]{1.2cm}{\ssize 
$\phi_0=\varphii$}} \\
&=&  
\left(\prod_{k=1}^{n-1}\prod_{\v{x}}\int\d\phi_k(\v{x})\right)
\left(\prod_{k=0}^{n-1}\prod_{\v{x}}\int\d\pi_k(\v{x})\right)
\left(\prod_{\v{x}}\sqrt{\frac{a^{d-1}}{2\pi\hbar}}\right)^{2n}\,\times 
\\
&&\times\,
\exp\left\{\frac{1}{\hbar}\sum_{k=0}^{n-1}a
\sum_{\v{x}}a^{d-1}\left[\i\frac{\phi_{k+1}(\v{x})
-\phi_k(\v{x})}{a}\pi_k(\v{x}) 
- \frac{1}{2}\pi_k^2(\v{x}) - \frac{1}{2}(\v{\nabla}\phi_k)^2(\v{x}) 
- 
\frac{1}{2}m^2\phi_k^2(\v{x})\right]\right\}_{\Big|\parbox{1.2cm}{\ssize 
$\phi_0=\varphii\,,$ \\ $\phi_n=\varphif$}}\,.
\eean
Integration over the momenta yields the path integral
\bea
\lefteqn{\left(\prod_{k=1}^{n-1}\prod_{\v{x}}\int\d\phi_k(\v{x})\right)
\left(\prod_{\v{x}}\left(\frac{a^{d-1}}{2\pi\hbar}\right)^n\left(\frac{2\pi\hbar}{a^d}\right)^{n/2}\right)\,\times} 
\nonumber \\
&& 
\times\,\exp\left\{-\frac{1}{\hbar}\sum_{k=0}^{n-1}a\sum_{\v{x}}a^{d-1}\left[\frac{1}{2}\left(\frac{\phi_{k+1}(\v{x})-\phi_k(\v{x})}{a}\right)^2 
+ \frac{1}{2}(\v{\nabla}\phi_k)^2(\v{x}) + 
\frac{1}{2}m^2\phi_k^2(\v{x})\right]
\right\}_{\Big|\parbox{1.2cm}{\ssize 
$\phi_0=\varphii\,,$ \\ $\phi_n=\varphif$}}\,.
\eea
In the zeroth and $n$th layer, $\phi$ is fixed to the initial and 
final values $\varphii$ and $\varphif$ respectively, while it is 
integrated over from layer 1 to $n-1$, weighted by the exponentiated 
action.

We can make this formula more symmetric with respect to the 
boundaries $\ti$ and $\tf$, if we add potential terms to the $n$th 
layer, writing
\bea
\lefteqn{\left(\prod_{k=1}^{n-1}\prod_{\v{x}}
\int\d\phi_k(\v{x})\right)
\left(\prod_{\v{x}}\left(\frac{a^{d-2}}{2\pi\hbar}
\right)^{n/2}\right)\,\times}
\nonumber \\
&& 
\times\,\exp\left\{-\frac{1}{\hbar}\sum_{\v{x}}a^d
\left[\sum_{k=0}^{n-1}\frac{1}{2}\left(\frac{\phi_{k+1}
(\v{x})-\phi_k(\v{x})}{a}\right)^2 
+ \sum_{k=0}^n\left(\frac{1}{2}(\v{\nabla}\phi_k)^2(\v{x}) + 
\frac{1}{2} m^2\phi_k^2(\v{x})\right)\right]
\right\}_{\Big|
\parbox{1.2cm}{\ssize $\phi_0=\varphii\,,$ \\ $\phi_n=\varphif$}}
\,.
\label{potentialaddedtonthlayer}
\eea
Clearly, such a change does not affect the continuum limit. We also 
rewrite the normalization factors: in \eq{potentialaddedtonthlayer}, 
there are $n$ factors of 
\be
\label{normalizationfactor}
C_a := \sqrt{\frac{a^{d-2}}{2\pi\hbar}}
\ee
for each $\v{x}\in S_a$. We express this in a more geometric fashion 
by attributing a factor $C_a$ to every spacetime point $x = 
(\v{x},\ti+ka)$ for which $\phi_k(\v{x})$ is integrated over, and by 
associating a factor $\sqrt{C_a}$ to each point in the initial and 
final layer:
\bea
\lefteqn{W_a[\varphif,\tf;\varphii,\ti] := 
\left(\prod_{\v{x}\in S_a}\sqrt{C_a}\right)^2
\left(\prod_{k=1}^{n-1}\prod_{\v{x}\in S_a}\int 
C_a\,\d\phi_k(\v{x})\right)\,\times} 
\nonumber \\
&& \times\,\exp\left\{-\frac{1}{\hbar}\sum_{\v{x}\in 
S_a}a^d\left[\sum_{k=0}^{n-1}\frac{1}{2}
\left(\frac{\phi_{k+1}(\v{x})-\phi_k(\v{x})}{a}\right)^2 
+ \sum_{k=0}^n\left(\frac{1}{2}(\v{\nabla}\phi_k)^2(\v{x}) + 
\frac{1}{2}m^2\phi_k^2(\v{x})\right)\right]
\right\}_{\Big|\parbox{1.2cm}{\ssize 
$\phi_0=\varphii\,,$ \\ $\phi_n=\varphif$}}\,.
\label{factorsredistributed}
\eea
In \fig{diagramVfi}, this is represented diagrammatically for the 
case $d=2$: open points stand for an integration over the associated 
field variable and carry a factor $C_a$. Boundary points are solid 
and contribute a factor $\sqrt{C_a}$. For each point there is a mass 
term in the action, and each link between points gives a term with 
the corresponding lattice derivative. The dual lattice is drawn 
shaded. \psfrag{n}{$n$}\psfrag{0}{$0$}\psfrag{vdots}{$\vdots$}
\pic{diagramVfi}{Lattice diagram for path integral on $V_{\rm 
fi}$.}{5cm}{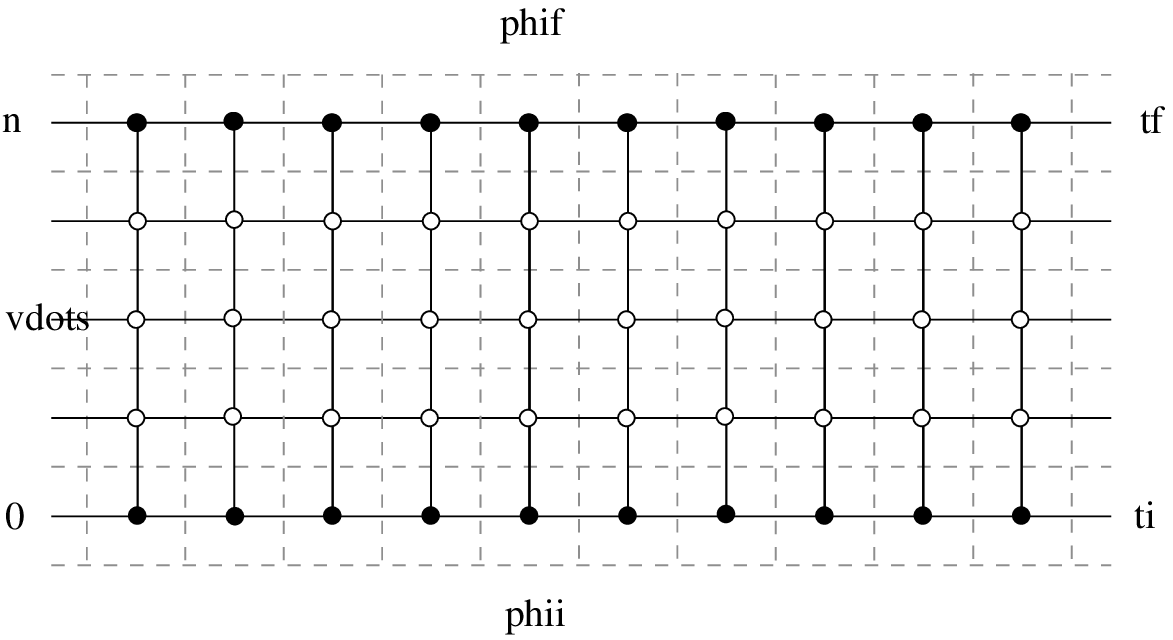}

\subsection{General Definition}

By applying the same rules to more complicated arrangements of 
points, we can define a path integral regularization for general 
volumes $V$.
Let $V\subset\bR^d$ be open and its boundary $\Sigma$ piecewise 
smooth. We use hypercubic lattices
$$L_a := \{x\in a\bZ^d\;|\;-aM\le |x_\mu|\le aM\,,\;\mu=1,\ldots,d\} 
$$ 
with lattice constant $a>0$ and edge length $2aM$, $M\in\bN$. 
$e_\mu$ is the unit vector in the $\mu$th direction. A lattice point 
$x$ and a direction $\mu$ define a link
$$l \equiv (x,\mu)\,.$$
The associated lattice gradient is
$$\nabla_l f \equiv \nabla_{\!\mu} f(x) := \frac{f(x+e_\mu) - 
f(x)}{a}\,.$$
Given a subset $P\subset L_a$, $l(P)$ denotes the set of links that 
connect points within $P$.
Let
$$\tilde{V}_a := L_a\cap V$$
be the intersection of $V$ with the lattice.
\psfrag{Sigma}{$\Sigma$}
\pic{diagramgeneralV}{Lattice diagram for a general volume 
$V$.}{5cm}{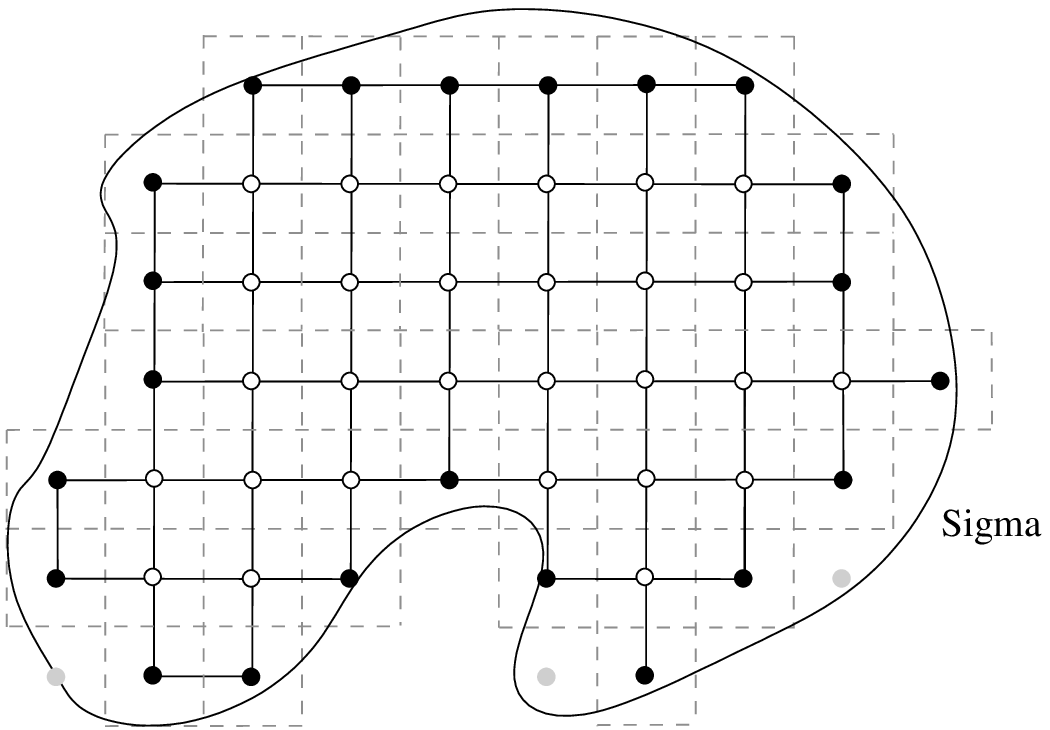}
The points of $\tilde{V}_a$ fall into three categories 
(\fig{diagramgeneralV}): we call a point {\it interior} if it has 
$2d$ links to points of $\tilde{V}_a$. If a point is linked to less 
than $2d$ points of $\tilde{V}_a$, but connected to at least one 
interior point, it is a {\it boundary point}. The remaining points of 
$\tilde{V}_a$ have only links to boundary points and we will not use 
them when representing the path integral on the lattice (they are 
drawn shaded in \fig{diagramgeneralV}). The set of relevant points is 
therefore
$$V_a := I_a\cup\Sigma_a\,,$$
where $I_a$ and $\Sigma_a$ denote the set of interior and boundary 
points respectively.

On the lattice, the path integral becomes a summation over scalar 
fields $\phi: V_a\to\bR$ on $V_a$. The action consists of 
contributions from links in $l(V_a)$ and points in $V_a$:
$$
S[\phi,V_a] := \sum_{l\in l(V_a)}a^d\,\frac{1}{2}(\nabla_l\phi)^2 + 
\sum_{x\in V_a}a^d\,\frac{1}{2}m^2\phi(x)
$$
Given a continuous boundary field $\varphi$ on $\Sigma$, one has to 
translate it into boundary data for $V_a$. We do so by defining the 
discrete boundary field
$$\varphi_a: \Sigma_a\to \bR\,,\quad \varphi_a(x) = \varphi({\rm 
pmd}_\Sigma(x))\,,$$
The function ${\rm pmd}_{\Sigma}$ (pmd stands for ``point of minimal
distance") returns a point on $\Sigma$ which has minimal distance to
$x$.  Now, we have all the necessary notation to give the regularized
form of the propagator $W[\varphi,V]$: we specify it as \be
\label{regularizedpropagatorV}
W_a[\varphi_a,V_a] := 
\left(\prod_{x\in\Sigma_a}\sqrt{C_a}\right)\left(\int\prod_{x\in 
I_a}C_a\,\d\phi(x)\right)\exp\left(-\frac{1}{\hbar}S[\phi,V_a]\right)
_{\Big|\phi|_{\Sigma_a} = \varphi_a}\,,
\ee 
with factors $C_a$ as in \eq{normalizationfactor}. The continuum 
propagator $W[\varphi,V]$ is then defined by the limit of vanishing 
lattice constant and infinite lattice size:
$$W[\varphi,V] := \lim_{a\to 0}\lim_{M\to\infty} 
W_a[\varphi_a,V_a]\,.$$
To simplify notation, we omit the  $\lim\limits_{M\to\infty}$--symbol 
in the remainder of the text. That is, the limit of infinite lattice size (for constant $a$) is implicit in all subsequent formulas. 
\vspace{0.5cm} 

We now make a number of unproven assumptions about the regularization
\eq{regularizedpropagatorV}:
\begin{description}
\item[(A1)] 
The propagator \eq{regularizedpropagatorV} has a continuum limit: 
that is, there is a non-trivial space $F(\Sigma)$ of boundary fields on $\Sigma$ such that for each $\varphi\in F(\Sigma)$ the limit 
$$
W[\varphi,V] := \lim_{a\to 0} W_a[\varphi_a,V_a]
$$
is well-defined.
\item[(A2)] 
$W$ reproduces the conventional propagator: for $V_{\rm fi} = 
\bR^{d-1}\times [\ti,\tf]$ and appropriate boundary conditions at 
spatial infinity, 
$$W[(\varphif,\varphii),V_{\rm fi}] = 
\b\varphif|\e^{-H(\tf-\ti)/\hbar}|\varphii\k\,.$$
\item[(A3)] $W[\varphi,V]$ is translation and rotation invariant: 
i.e.\ under an isometry $f:\bR^d\to\bR^d$,
$$W[\varphi\circ f^{-1},f(V)] = W[\varphi,V]\,.$$
\item[(A4)] There is a functional derivative $\ds\fdiff{}{\varphi}$ on 
$F(\Sigma)$ whose action on $W[\varphi,V]$ can be approximated as 
follows:
$$\sum_{x\in\Sigma_a} a^{d-1}\pdiff{^n 
W_a}{(a^{d-1}\varphi_a(x))^n}[\varphi_a,V_a]\quad\stackrel{a\to 
0}{\longrightarrow}\quad\int_\Sigma\d\Sigma(x)\fdiff{^n 
W}{\varphi(x)^n}[\varphi,V]\,.$$
\end{description}
\vspace{0.5cm}
To evaluate the path integral \eq{regularizedpropagatorV}, it is 
useful to arrange the field variables from each point in vectors 
$\phi$ and write the action as
\be
\label{action}
S[\phi,V_a] = \frac{1}{2}\phi\cdot B_a\,\phi + c_a\cdot\phi + d_a\,.
\ee
The boundary fields $\varphi_a$ are contained in the vectors $c_a$ 
and $d_a$ respectively.
The action is bounded from below, so for each $\varphi_a$, there is 
at least one solution $\phi_{\rm cl}$ of the Euclidean equations of 
motion
$$\pdiff{S}{\phi}[\varphi_a,V_a] = B_a\,\phi + c_a = 0\,.$$
If $B_a$ is non-degenerate, the solution is unique and one can define 
the Hamilton function
$$S[\varphi_a,V_a] := S[\phi_{\rm cl},V_a]$$
for the discrete Euclidean system. We assume, in fact, that
\vspace{0.5cm}
\begin{description}
\item[(A5)] The matrix $B_a$ is non-degenerate and the Hamilton 
 function $S[\varphi_a,V_a]$ is analytic in $\varphi_a$.
\end{description}
\vspace{0.5cm}
The change of variables $\xi := \phi - \phi_{\rm cl}$ renders the 
action \eq{action} quadratic:
$$ S[\xi,V_a] = \frac{1}{2}\xi\cdot B_a\,\xi + S[\varphi_a,V_a]$$
The integral \eq{regularizedpropagatorV} becomes Gaussian and gives
\be
\label{pathintegralevaluated}
W_a[\varphi_a,V_a] = 
\left(\prod_{x\in\Sigma_a}\sqrt{C_a}\right)\left(\int\prod_{x\in 
I_a}\sqrt{2\pi}\,C_a\right)\frac{1}{\sqrt{\det 
B_a}}\,\exp\left(-\frac{1}{\hbar}S[\varphi_a,V_a]\right)\,.
\ee
Therefore, by {\bf (A5)}, the regularized kernel $W_a[\varphi_a,V_a]$ 
must be analytic in $\varphi_a$, which will be used in section 
\ref{Discrete_Schroedinger_Equation} when deriving the Schr\"odinger 
equation. \\

\noindent {\it Remark:} In {\bf (A1)} and {\bf (A5)} we have formulated the continuum limit in terms of pointwise convergence, i.e.\ by separate convergence for each boundary field $\varphi$ in $F(\Sigma)$. According to \eq{pathintegralevaluated}, the field dependence of the regularized kernel resides only in the Hamilton function $S[\varphi_a,V_a]$. The latter converges against the continuum function $S[\varphi,V]$, which is defined pointwise. Thus, it is plausible to assume that the continuum propagator $W[\varphi,V]$, too, is a pointwise function on $F(\Sigma)$. 
When further developing the formalism, pointwise convergence is likely to be replaced by convergence in a Hilbert space norm or other measures which only distinguish between equivalence classes of boundary fields. For the purpose of this article, however, it is sufficient and simplifies notation if we use convergence on single fields.

\section{Generalized Schr\"odinger Equation}
\label{Generalized_Schroedinger_Equation}

The propagator $W$ depends on a spacetime region $V$ and a field 
$\varphi$ specified on the boundary $\Sigma$. Thus, as for the 
Hamilton  function in section 
\ref{Generalized_Hamilton_Jacobi_Equation}, one can define a 
deformation derivative: using the same notation as there, we set
$$L_N W[\varphi,V] := \lim_{s\to 0} \frac{W[\varphi^s,V^s] - 
W[\varphi,V]}{s}\,.$$
In this section, we derive that
\be
\label{generalizedSchroedingerequation}
\hbar L_N W[\varphi,V] = \left(-H_N[\varphi,\hbar\fdiff{}{\varphi},V] 
+ P_N[\varphi,\hbar\fdiff{}{\varphi},V]\right)W[\varphi,V]\,,
\ee
where 
\bean
H_N[\varphi,\pi,V] &:=& 
\int_\Sigma\d\Sigma\;N_\perp\!\left(-\frac{1}{2}\pi^2 +
\frac{1}{2}(\naSig\varphi)^2 + \frac{1}{2}m^2\varphi^2\right)\,, \\
P_N[\varphi,\pi,V] &:=&
-\int_\Sigma\d\Sigma\,N_\parallel\cdot\naSig\varphi\,\pi\,.
\eean
When $V = \bR^{d-1}\times [\ti,\tf]$ and $N_\perp|_{\tf} = 1$, 
$N_\perp|_{\ti} = 0$,
$N_\parallel = 0$, this yields the ordinary Euclidean Schr\"odinger 
equation 
$$\left(\hbar\pdiff{}{\tf} +
H[\varphif,\hbar\fdiff{}{\varphif}]\right)W[\varphif,\tf;\varphii,\ti] 
=
0\,.$$
The strategy of the derivation: using assumption {\bf (A5)} and
rotation invariance {\bf (A3)}, we show that the regularized
propagator satisfies a lattice version of equation
\eq{generalizedSchroedingerequation} when $V$ is deformed along flat
parts of its boundary.  The central step is analogous to the
calculation Feynman used when deriving the Schr\"odinger equation from
the path integral of a point particle \cite{Feynman} (see also chap.\
4, \cite{Schulman}).  Due to {\bf (A1)} and {\bf (A4)}, the discrete
equations have the continuum limit
\eq{generalizedSchroedingerequation}.  To cover also the case, when
deformations are applied to curved sections of $\Sigma$, we
approximate $\Sigma$ by a triangulation, apply
\eq{generalizedSchroedingerequation} to each triangle and let the
fineteness of the triangulation go to zero.

For simplicity, the argument is formulated for {\it bounded} volumes 
below. The generalization to infinitely extended $V$ is 
straightforward.
 
\subsection{Discrete Schr\"odinger Equation}
\label{Discrete_Schroedinger_Equation}

\psfrag{Sigmaz}{$\Sigma_0$} \psfrag{Sigmaup}{$\Sigma'_1$}
\psfrag{Sigmau}{$\Sigma_1$}\psfrag{n}{$n$}\psfrag{Ha}{$H_a$}
\pic{additionoflayer}{Addition of a single
layer.}{4cm}{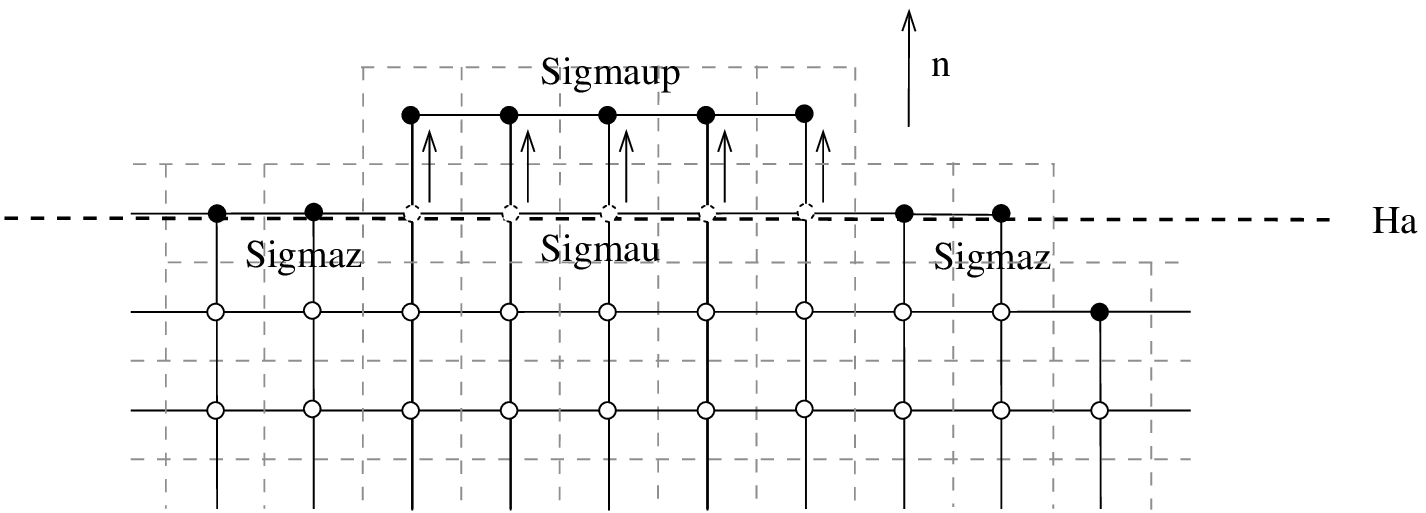} Consider a lattice diagram in which
part of $\Sigma_a$ coincides with a hypersurface $H_a$ of the lattice
$L_a$.  Let $n$ denote the normal vector of $H_a$.  The simplest way
of modifying such a diagram is to add a $(d-1)$-dimensional layer of
points along $H_a\cap\Sigma_a$ (see \fig{additionoflayer}).  The old
boundary points adjacent to the layer become interior points.  We
describe this by a lapse function $N_a: \Sigma_a\to\{0,1\}$ which
indicates for any given point of the boundary if a new point will be
linked to it or not.  Then, the function
 $$\sigma_a:
\Sigma_a\to L_a\,,\;\; x\mapsto x + a N_a(x)\,n$$
 is the discrete flow
associated to the deformation of the boundary $\Sigma_a$.
 Define the new
diagram and its boundary by
$$V'_a :=
V_a\cup\sigma_a(\Sigma_a)\,,\qquad\Sigma'_a := \sigma_a(\Sigma_a)\,.$$
The set
$$\Sigma_1 := N_a^{-1}(1)$$
is the part of $\Sigma_a$ which is moved and becomes $\Sigma'_1 := 
\sigma_a(\Sigma_1)$, while
$$\Sigma_0 := N_a^{-1}(0)$$
remains unchanged. As in the continuous case, we choose the new
boundary field to be the pull-forward of the old one, that is,
$$\varphi'_a := \varphi_a\circ\sigma_a^{-1}\,.$$
The resulting path integral is 
\bean
\lefteqn{W_a[\varphi'_a,V'_a] = 
\left(\prod_{x\in\Sigma_0\cup\Sigma'_1}\sqrt{C_a}\right)
\left(\prod_{x\in
I_a\cup\Sigma_1}\int C_a\,\d\phi(x)\right)\,\times} \\
&&
\times\,\exp\left\{
-\frac{a^d}{\hbar}\left[\sum_{x\in\Sigma_1}\frac{1}{2}
\left(\frac{\varphi'_a(\sigma_a(x)) 
- \phi(x)}{a}\right)^2 + \sum_{l\in 
l(\Sigma'_1)}\frac{1}{2}(\nabla_l\varphi'_a)^2 + \sum_{x\in 
\Sigma'_1}\frac{1}{2}m^2\varphi_a'{}^2(x)\right]\right. \\
&& \hspace{1.4cm}{}-\frac{1}{\hbar}S[\phi,V_a]\Bigg\}
_{\Big|{\parbox{1.8cm}{
 \ssize $\phi|_{\Sigma_0}=\varphi_a|_{\Sigma_0}$}}}
\eean
(Recall that $l(\Sigma'_1)$ is the set of links between points of 
$\Sigma'_1$.)
The same can also be written as a convolution with the original 
propagator:
\bean
\lefteqn{\left(\prod_{x\in\Sigma_1}\int 
C_a\,\d\phi(x)\right)\exp\left\{-\frac{a^d}{\hbar}
\left[\sum_{x\in\Sigma_1}\frac{1}{2}\left(\frac{\varphi_a(x) 
- \phi(x)}{a}\right)^2\right.\right.} \\
&&\hspace{3.4cm}\left.\left.{}+ \sum_{l\in 
l(\Sigma_1)}\frac{1}{2}(\nabla_l\varphi_a)^2 + 
\sum_{x\in\Sigma_1}\frac{1}{2}m^2\varphi_a^2(x)\right]\right\}
\,W_a[(\varphi_a|_{\Sigma_0},\phi),V_a]
\eean
Following Feynman's derivation of the Schr\"odinger equation, we 
introduce new variables
$$\xi(x) := \sqrt{\frac{a^{d-2}}{\hbar}}(\phi(x) - 
\varphi_a(x))\,,\quad x\in\Sigma_1$$
and get
\bean
W_a[\varphi'_a,V'_a] &=& 
\Bigg(\prod_{x\in\Sigma_1}
\int\overbrace{C_a\sqrt{\frac{\hbar}{a^{d-2}}}}^{=1/\sqrt{2\pi}}\,
\d\xi(x)\Bigg)
\exp\left\{-\frac{1}{2}\sum_{x\in\Sigma_1}\frac{\xi^2(x)}{a}
- \frac{a^d}{\hbar}\left[\sum_{l\in 
l(\Sigma_1)}\frac{1}{2}(\nabla_l\varphi_a)^2 + 
\sum_{x\in\Sigma_1}\frac{1}{2}m^2\varphi_a^2(x)\right]\right\} \\
&& \hspace{5mm}
\times\:\:W_a\left[\left(\varphi_a|_{\Sigma_0},\varphi_a|_{\Sigma_1}
+\sqrt{\hbar/a^{d-2}}\,\xi\right)\,,\,V_a\right]
\eean
Next we apply Laplace's method to obtain an asymptotic expansion of 
this expression (see e.g.\ chap.\ 11, \cite{Schulman}): the dominant 
contribution to the Gaussian integral comes from an $a$-dependent 
interval $[-\varepsilon_a,\varepsilon_a]^{|\Sigma_1|}$ around $\xi = 
0$. ($|\Sigma_1|$ denotes the number of points in $\Sigma_1$.) The 
integral outside is exponentially damped for $a\to 0$ and neglected. 
Within the interval, one can Taylor expand $W_a$ in $\xi$ and reverse 
the order of integration and Taylor expansion. To evaluate the 
integration for each term, the integration range is extended back to 
its full size: this introduces an error in each term of the sum and 
convergence is lost, but the expansion is still valid asymptotically 
for $a\to 0$.

For the integrations and estimates, we use the following formulas:
\bea
\label{Gaussianintegration}
\int\limits_{-\infty}^{\infty}\d y\,y^n\,\e^{-y^2/2}
&=& \Bigg\{
\parbox{7cm}{
$$
\begin{array}{lcl}
(n-1)(n-3)\cdots 3\cdot 1\cdot\sqrt{2\pi} & , & \mbox{$n\ge 0$ and 
even}\,, \\
&& \\
0 & , & \mbox{$n$ odd}\,, 
\end{array}
$$
}
\\
\label{exponentiallydamped}
\int\limits_{\pm \varepsilon_a}^{\pm \infty}\d y\,y^n\,\e^{-y^2/2} 
&=& O(\varepsilon^{n-1}_a\,\e^{-\varepsilon^2_a/2})\quad\mbox{as 
$a\to 0$\,.}
\eea
We set $\epsilon_a = 1/a$. Consider first the integral outside the 
chosen interval:
\bea
\lefteqn{
\left(\prod_{x\in\Sigma_1}\frac{1}{\sqrt{2\pi}}
\int\limits_{\bR\backslash 
[-1/a,1/a]}\d\xi(x)\right) 
\exp\left(-\frac{1}{2}\sum_{x\in\Sigma_1}\frac{\xi^2(x)}{a}\right)
\exp\left\{-\frac{a^d}{\hbar}\left[\sum_{l\in 
l(\Sigma_1)}\frac{1}{2}(\nabla_l\varphi_a)^2 + 
\sum_{x\in\Sigma_1}\frac{1}{2}m^2\varphi_a^2(x)\right]\right\} 
} \nonumber \\
&& \label{integraloutsideinterval}
\times\:\:W_a\left[\left(\varphi_a|_{\Sigma_0},
\varphi_a|_{\Sigma_1}+\sqrt{\hbar/a^{d-2}}\,\xi\right)\,,\,V_a\right] 
\hspace{8.8cm}
\eea
The second exponent vanishes in the continuum limit. For $W_a$, we 
employ formula \eq{pathintegralevaluated} and replace 
$\exp(-S[\ldots,V_a])$ by 1, as the action is positive. The 
determinant and $C_a$-factors are together of order $O(1)$, since, by 
assumption, \eq{pathintegralevaluated} approaches a finite continuum 
limit. Thus, the modulus of \eq{integraloutsideinterval} is smaller 
than
$$
\left(\prod_{x\in\Sigma_1}\frac{1}{\sqrt{2\pi}}
\int\limits_{\bR\backslash 
[-1/a,1/a]}\d\xi(x)\,\e^{-\xi^2(x)/2}\right)\cdot 
O(1)\quad\stackrel{\eq{exponentiallydamped}}{\le}\quad 
O\left(\frac{1}{a}\,\e^{\ts 
-\frac{|\Sigma_1|}{2a^2}}\right)\quad\mbox{as $a\to 0\,.$}
$$
In the integral over $[-1/a,1/a\,]^{|\Sigma_1|}$, we Taylor expand 
$W_a$ in $\xi$: 
\bean
\lefteqn{\left(\prod_{x\in\Sigma_1}\frac{1}{\sqrt{2\pi}}
\int_{-1/a}^{1/a}\d\xi(x)\right)
\exp\left(-\frac{1}{2}\sum_{x\in\Sigma_1}
\frac{\xi^2(x)}{a}\right)\exp\left\{
-\frac{a^d}{\hbar}\left[\sum_{l\in 
l(\Sigma_1)}\frac{1}{2}(\nabla_l\varphi_a)^2 + 
\sum_{x\in\Sigma_1}\frac{1}{2}m^2\varphi_a^2(x)\right]\right\}} \\
&&\times\,
\left(W_a[\varphi_a,V_a] + 
\sum_{x\in\Sigma_1}\pdiff{W_a}{\varphi_a(x)}
\sqrt{\frac{\hbar}{a^{d-2}}}\,\xi(x) 
+ \frac{1}{2}\sum_{x,y\in\Sigma_1}\pdiff{^2W_a}
{\varphi_a(x)\partial\varphi_a(y)}\,\frac{\hbar}{a^{d-2}}\,\xi(x)\xi(y) 
+ \ldots\right) \nonumber
\eean
By assumption {\bf (A5)}, $W_a$ is analytic in the field variable, so 
the Taylor expansion converges uniformly and we are allowed to 
integrate each term of the series separately. We also set the limits 
of integration back to plus and minus infinity. This does not affect 
the {\it asymptotic} property of the series, since for each term the 
resulting error is only exponentially small: for example , the linear 
term yields
$$
\left|\left(\prod_{x\in\Sigma_1}\frac{1}{\sqrt{2\pi}}\int_{-1/a}^{1/a}\d\xi(x)\right)\e^{-\xi^2(x)/2}\sum_{x\in\Sigma_1}a^{d-1}\pdiff{W_a}{(a^{d-1}\varphi_a(x))}\,\sqrt{\frac{\hbar}{a^{d-2}}}\,\xi(x)\right| 
\quad\le\quad 
O\left(\frac{|\Sigma_1|}{\sqrt{a^{d-2}}}\,\e^{-\ts\frac{1}{2a^2}}\right)\,,
$$
because of \eq{exponentiallydamped} and {\bf (A4)}. Then, we can use 
equation \eq{Gaussianintegration} to do the Gaussian integration in 
each term of the asymptotic series. Each integration, that is, each 
point $x\in\Sigma_1$, leaves an overall factor $\sqrt{2\pi}$. Terms 
with an uneven number of $\xi$ variables (of the same point) vanish. 
We obtain
\bean
\lefteqn{W_a[\varphi'_a,V'_a] \:\: \sim \:\: 
\exp\left\{-\frac{a^d}{\hbar}\left[\sum_{l\in 
l(\Sigma_1)}\frac{1}{2}(\nabla_l\varphi_a)^2 + 
\sum_{x\in\Sigma_1}\frac{1}{2}m^2\varphi_a^2(x)\right]\right\}} \\
&&\times\,
\left(W_a[\varphi_a,V_a] + 
\sum_{x\in\Sigma_1}\frac{\hbar}{2}
\pdiff{^2W_a}{\varphi_a(x)^2}\cdot\frac{1}{a^{d-2}} 
+ \sum_{n=2}^\infty \sum_{x\in\Sigma_1}c(n)\pdiff{^{2n} 
W_a}{\varphi_a(x)^{2n}}\cdot\frac{1}{(a^{d-2})^n}\right)\,,
\eean
where the $c(n)$'s are numerical coefficients. If we write 
$\varphi_a$ as the pull-back $\varphi'_a\circ\sigma_a \equiv 
\sigma_a^*\varphi'_a$ of $\varphi'_a$ and use also the lapse function 
$N_a$, the final result becomes
\bea
\lefteqn{W_a[\varphi'_a,V'_a] \:\: \sim \:\: \exp\left\{
-\frac{1}{\hbar}\left[\sum_{l\in 
l(\Sigma_a)}a^{d-1}\,aN_a\,\frac{1}{2}
(\nabla_l\sigma_a^*\varphi'_a)^2 
+ \sum_{x\in\Sigma_a}a^{d-1}\,aN_a\frac{1}{2}
m^2(\sigma_a^*\varphi'_a)^2(x)\right]\right\}} 
\nonumber \\
&& \times\,\left\{W_a[\sigma_a^*\varphi'_a,V_a] + \sum_{x\in\Sigma_a} 
a^{d-1}\,aN_a\left(\frac{\hbar}{2}
\pdiff{^2W_a}{(a^{d-1}\varphi_a(x))^2}
[\sigma_a^*\varphi'_a,V_a]\right.\right. 
\nonumber \\
&& \left.\left.\hspace{0.7cm}{}+ \sum_{n=2}^\infty 
a^{d(n-1)}\,c(n)\pdiff{^{2n} 
W_a}{(a^{d-1}\varphi_a(x))^{2n}}[\sigma_a^*\varphi'_a,V_a]
\right)\right\}\,.\hspace{5cm}
\label{singlelayer}
\eea

\subsubsection*{Iteration}

Suppose now that the deformed set $V'_a$ does not arise from the 
addition of a single layer, but from a continuous deformation $V^s$ 
of the orginial volume $V$. That is, we want to compare 
$W_a[\varphi^s_a,V^s_a]$ against $W_a[\varphi_a,V_a]$. To make the 
calculation tractable, we require that the vector field $N$ vanishes 
outside a neighbourhood $U$ of a boundary point $x\in\Sigma$, and 
that within $U$ the boundary $\Sigma$ is flat. Denote this part of 
$\Sigma$ by $\Sigma_U := \Sigma\cap U$.

\psfrag{sigmaa}{$\sigma_a$}\psfrag{pmd}{${\rm 
pmd}_{\Sigma^s}$}\psfrag{sigmasmu}{$\sigma_s^{-1}$}
\pic{diagramVs}{Diagram for $V^s_a$.}{5cm}{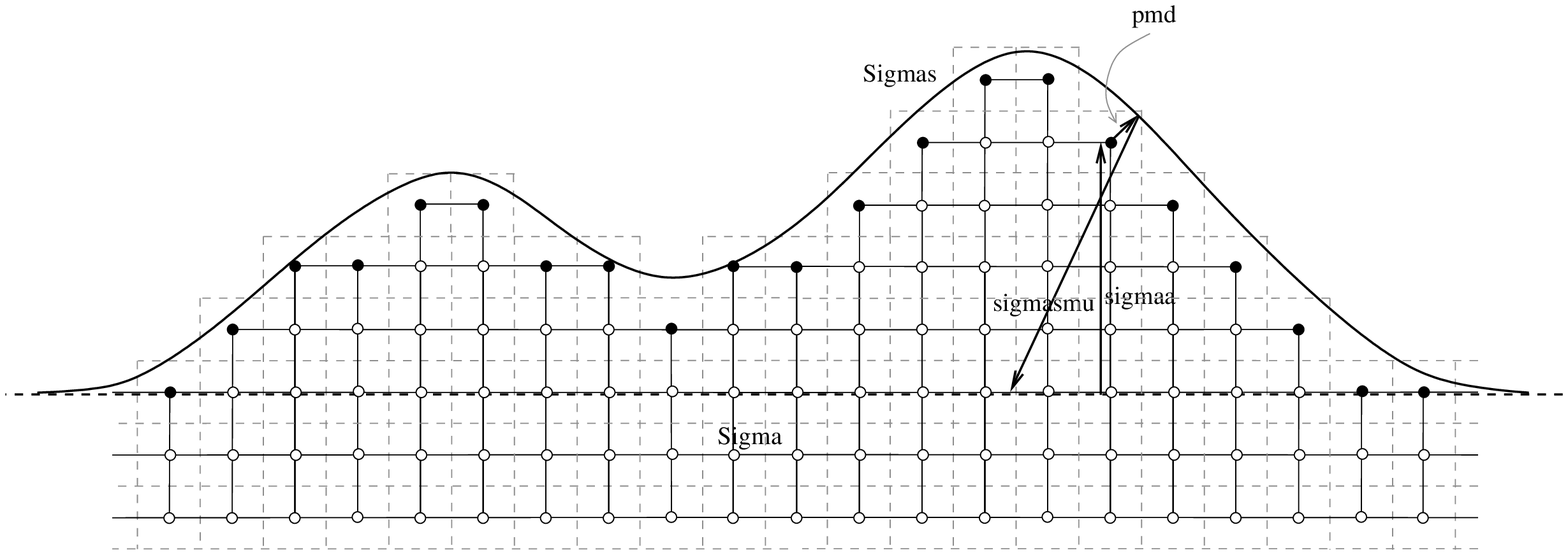}
By rotation and translation invariance ({\bf (A3)}), we can orient 
$V$ such that $\Sigma_U$ coincides with a hyperplane of the lattice 
$L_a$. Let us begin by considering the case where the lapse $N_\perp$ 
is positive, that is, $V\subset V^s$. For small enough $s$, the 
typical diagram for $W_a[\varphi^s_a,V^s_a]$ looks like 
\fig{diagramVs} (or its higher-dimensional equivalent), where along 
the normal direction $n$ each point of $\Sigma_a$ is in one-to-one 
correspondence with a point of $\Sigma^s_a$. (Note that in the 
limit $s\to 0$, the slope of $\Sigma^s$ against $\Sigma$ becomes 
arbitrarily small.) The new boundary $\Sigma^s_a$ can be built from 
$\Sigma_a$ by repeatedly adding single layers as described 
previously. Thus, we can iterate formula \eq{singlelayer} to obtain a 
relation between $W_a[\varphi^s_a,V^s_a]$ and 
$W_a[\sigma_a^*\varphi^s_a,V_a]$ where now, $\sigma_a$ is the 
concatenation of all single-step flows. When collecting the various 
terms of the iteration, the lapse functions for each step add up to 
the {\it total} lapse function $N_a$. We order the result in powers 
of $a N_a$ and $a$:
\bea
W_a[\varphi^s_a,V^s_a] &=& W_a[\sigma_a^*\varphi^s_a,V_a] + 
\sum_{x\in\Sigma_a} 
a^{d-1}\,aN_a\frac{\hbar}{2}\pdiff{^2W_a}{(a^{d-1}
\varphi_a(x))^2}[\sigma_a^*\varphi^s_a,V_a] 
\nonumber \\
&& {}-\frac{1}{\hbar}\left[\sum_{l\in 
l(\Sigma_a)}a^{d-1}\,aN_a\,\frac{1}{2}
(\nabla_l\sigma_a^*\varphi^s_a)^2 
+ 
\sum_{x\in\Sigma_a}a^{d-1}\,aN_a\frac{1}{2}m^2
(\sigma_a^*\varphi^s_a)^2(x)\right] 
\nonumber \\  
&& {}+ O(a^d(aN_a)) + O((aN_a)^2)\qquad\mbox{as $a,s\to 0\,.$}
\label{afteriteration}
\eea
Note that the displacement vector $aN_a$ approaches $sN_\perp$ when 
both $a$ and $s$ become small, i.e.
\be
\label{sNperp}
aN_a = sN_\perp + O(s^2) + O(a)\,.
\ee
If $N$ is normal to $\Sigma_U$, the discrete flow $\sigma_a$ 
approximates the continuous one, $\sigma_s$, and
\bean
\sigma_a^*\varphi^s_a &=& \varphi^s_a\circ\sigma_a = 
\varphi^s\circ{\rm pmd}_{\Sigma_s}\circ\sigma_a \\
&=& \varphi\circ\sigma_s^{-1}\circ{\rm pmd}_{\Sigma_s}\circ\sigma_a \\
&=& \varphi_a + O(a)\,.
\eean
In general, $N$ has also a tangential component, so
\be
\sigma_a^*\varphi^s_a = \varphi_a - 
sN^\mu_\parallel\nabla_{\!\mu}\varphi_a  + O(s^2) + O(a)\,,
\label{tangentialshift}
\ee
as can be seen from the arrow diagram in \fig{diagramVs}. 
Plugging \eq{sNperp} and \eq{tangentialshift} into 
\eq{afteriteration}, one arrives at a regularized form of the 
Euclidean Schr\"odinger equation: 
\be
\label{discreteSchroedingerequation}
\frac{W_a[\varphi^s_a,V^s_a] - W_a[\varphi_a,V_a]}{s} = \hat{O}_a 
W_a[\varphi_a,V_a] + O(s) + O(a) + O(a/s)\,,
\ee
where $\hat{O}_a$ is the operator
\pic{posneglapse}{Lapse with positive and negative 
sign.}{4cm}{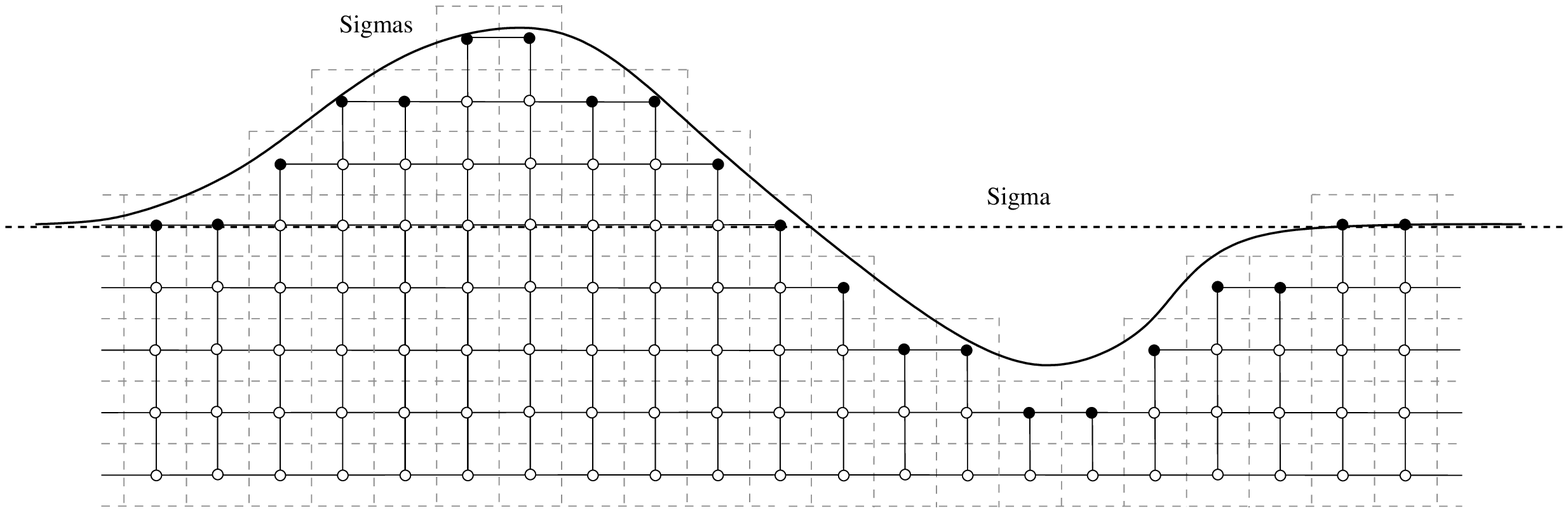}
\bea
\hat{O}_a &:=& -\frac{1}{\hbar}\sum_{x\in\Sigma_a} 
a^{d-1}\left[N_\perp(x)\left(-\frac{\hbar^2}{2}
\pdiff{^2}{(a^{d-1}\varphi_a(x))^2} 
+ \frac{1}{2}m^2\varphi_a^2(x)\right) + 
N^\mu_\parallel(x)\nabla_{\!\mu}\varphi_a(x)\,
\hbar\pdiff{}{(a^{d-1}\varphi_a(x))}\right] 
\nonumber \\
&& \hspace{2.5cm}{} - \frac{1}{\hbar}\sum_{l\in 
l(\Sigma_a)}a^{d-1}N_\perp(x)\,\frac{1}{2}(\nabla_l\varphi_a)^2
\label{regularizedO}
\eea
An analogous argument applies to the case of negative lapse 
$N_\perp$. For mixed diagrams as in \fig{posneglapse}, both types of 
calculations can be combined to give 
\eq{discreteSchroedingerequation} for lapses of arbitrary sign.


\subsection{Continuous Schr\"odinger Equation}
\label{Continuous_Schroedinger_Equation}

Choose $N$ as before, i.e.\ with support on a neighbourhood $U$ of 
$x\in\Sigma$ where $\Sigma\cap U$ is flat. We want to show that
\be
\label{continuousSchroedingerequation}
L_NW[\varphi,V] = \lim_{s\to 0} \frac{W[\varphi^s,V^s] - 
W[\varphi,V]}{s} = \hat{O}W[\varphi,V]\,,
\ee
where
\bean
\hat{O} &:=&
-\frac{1}{\hbar}\int_\Sigma\d\Sigma(x)
\left[N_\perp(x)\left(-\frac{\hbar^2}{2}\fdiff{^2}{\varphi(x)^2} 
+ \frac{1}{2}(\naSig\varphi)^2(x) + 
\frac{1}{2}m^2\varphi^2(x)\right)\right. \\
 &&
\hspace{2.5cm}{}+N_\parallel(x)\cdot\naSig\varphi(x)\,
\hbar\fdiff{}{\varphi(x)}\Bigg]\,. 
\eean
Stated more explicitly, this means that for any $\epsilon > 0$ there 
is an
$s_0>0$ such that
 \be
\label{mathematicaldefinition}
\left|\frac{W[\varphi^s,V^s] - W[\varphi,V]}{s} - 
\hat{O}W[\varphi,V]\right|\;\;\le\;\;\epsilon\qquad\mbox{for all 
$s<s_0$\,.}
\ee
To obtain an upper estimate on the left-hand side, we insert 
regularized propagators and operators in a suitable way, and then 
apply the triangle inequality:
\bean
\setlength{\jot}{2mm}
\mbox{{\sc lhs} of \eq{mathematicaldefinition}} &=& 
\left|\frac{1}{s}\Big(
W[\varphi^s,V^s] - W_a[\varphi^s_a,V^s_a] + W_a[\varphi^s_a,V^s_a] - 
W_a[\varphi_a,V_a] + W_a[\varphi_a,V_a] - W[\varphi,V]
\Big)\right. \\
&& {}- \hat{O}_aW_a[\varphi_a,V_a] + \hat{O}_aW_a[\varphi_a,V_a] - 
\hat{O}W[\varphi,V]\bigg| \\
&\le&  
\frac{1}{s}\Big|W[\varphi^s,V^s] - W_a[\varphi^s_a,V^s_a]\Big|
+ \frac{1}{s}\Big|W[\varphi,V] - W_a[\varphi_a,V_a]\Big| \\ 
&& {}+ \Big|\hat{O}W[\varphi,V] - \hat{O}_aW_a[\varphi_a,V_a]\Big| \\
&& {}+ \left|\frac{W_a[\varphi^s_a,V^s_a] - W_a[\varphi_a,V_a]}{s}
- \hat{O}_aW_a[\varphi_a,V_a]\right|\,.
\setlength{\jot}{0mm}
\eean
By assumption {\bf (A1)} (existence of the continuum limit), the 
first two terms become smaller than $\epsilon/4$ when the lattice 
constant $a$ is smaller than some $a_s>0$. The partial derivatives 
and potential terms in \eq{regularizedO} approach their continuum 
analogues as $a\to 0$, so there is also an $a_0>0$ such that
$$
\Big|\hat{O}W[\varphi,V] - \hat{O}_aW_a[\varphi_a,V_a]\Big| < 
\frac{\epsilon}{4}\qquad\mbox{for all $a<a_0$\,.}
$$
The regularized Schr\"odinger equation tells us that for $s$ smaller 
than some $s_0$, there is an $a'_s>0$ such that
$$\left|\frac{W_a[\varphi^s_a,V^s_a] - W_a[\varphi_a,V_a]}{s}
- \hat{O}_aW_a[\varphi_a,V_a]\right| < 
\frac{\epsilon}{4}\qquad\mbox{for all $a<a'_s$\,.}$$
Thus, for any $s<s_0$, we can choose $a < \min\{a_s,a_0,a'_s\}$ and 
the left-hand side of \eq{mathematicaldefinition} must be smaller 
than $\epsilon$. \qed

\subsection{Curved Boundaries}
\label{Curved_Boundaries}

As it is based on the lattice equation 
\eq{discreteSchroedingerequation}, the previous derivation applies 
only when flat sections of the boundary $\Sigma$ are deformed. We do 
not know how to extend the lattice calculation to the case where both 
initial and deformed surface are curved. Below we give an argument 
which circumvents this difficulty, but requires additional 
assumptions. The idea is to approximate the curved boundary by a 
triangulation, apply the variation to each of the flat triangles and 
add up the contributions.

Let $T_\delta$ be a triangulation of $\Sigma$ with fineness $\delta$: 
that is, when two 0-simplices are connected by a 1-simplex, their 
metric distance is at most $\delta$. Let $\{\Sigma_\alpha\}$ denote 
the set of $(d-1)$-simplices $\Sigma_\alpha\subset\Sigma$ of the 
triangulation. The corner points of each such simplex $\Sigma_\alpha$ 
define a $(d-1)$-simplex in $\bR^d$ which we call 
$\Sigma_{\Delta_\alpha}$. The hypersurface $\Sigma_{\Delta} := 
\cup_\alpha \Sigma_{\Delta_\alpha}$ approximates $\Sigma$ and 
encloses the volume $V_\Delta$. We can view $V$ as a deformation of 
$V_\Delta$ and find a flow 
$$\rho: \bR\times\bR^d \to \bR\,,\quad (t,x)\mapsto 
\rho(t,x)\equiv\rho_t(x)$$
such that $\rho_1(V_\Delta) = V$ and $\rho_1(\Sigma_{\Delta_\alpha}) 
= \Sigma_\alpha$.
We equip $\Sigma_\Delta$ with the boundary field $\varphi_\Delta := 
\rho_1^*\varphi = \varphi\circ\rho_1$, the pull-back of $\varphi$ 
under this flow.
Motivated by equation \eq{continuousSchroedingerequation} for flat 
surfaces, we assume that the difference between $W[\varphi,V]$ and 
$W[\varphi_\Delta,V_\Delta]$ is of the order of the volume difference 
between $V$ and $V_\Delta$:
\bea
W[\varphi,V] &=& W[\rho_{1*}\varphi_\Delta,\rho_1(V_\Delta)] 
\nonumber \\
&=& W[\varphi_\Delta,V_\Delta] + O(|V-V_\Delta|)\,.
\label{oforderofvolumedifference}
\eea
Next we introduce ``characteristic'' functions 
$\chi_\alpha:\bR^d\to\bR$ with the property that
$$
\renewcommand{\arraystretch}{1.2}
\begin{array}{llcl}
& \chi_\alpha(x) = 1 & \mbox{for} & x\in\Sigma_{\Delta_\alpha}\,, \\
& \chi_\alpha(x) = 0 & \mbox{for} & x\in\Sigma_{\Delta_\beta}\,,\; 
\alpha\neq \beta\,, \\
\mbox{and} & \multicolumn{3}{l}{\sum_\alpha\chi_\alpha(x) = 
1\;\;\mbox{for all}\;\;x\in\bR^d\,.}
\renewcommand{\arraystretch}{1}
\end{array}
$$
Using these functions, we can decompose the deformation field $N$ 
according to 
$$N = \sum_\alpha\chi_\alpha N \equiv \sum_\alpha N_\alpha\,.$$
Each component $N_\alpha$ is a discontinuous vector field and gives 
rise to a discontinuuous flow within $\bR^d$. Suppose that by a 
suitable limiting procedure, one can define $L_{N_\alpha}$ such that 
equation \eq{continuousSchroedingerequation} holds and
$$L_N = \sum_\alpha L_{N_\alpha}\,.$$
Then, equation \eq{oforderofvolumedifference} becomes
$$
L_NW[\varphi,V] = \sum_\alpha L_{N_\alpha} W[\varphi_\Delta,V_\Delta] 
+ O(|V-V_\Delta|)\,.
$$
By construction, the vector fields $N_\alpha$ are only nonzero on the 
flat simplices $\Sigma_{\Delta_\alpha}$. Therefore, our result for 
flat surfaces (equation \eq{continuousSchroedingerequation}) is 
applicable and yields
\bean
\lefteqn{L_NW[\varphi,V]} \\
&=& 
\sum_\alpha\left\{-\frac{1}{\hbar}\int_\Sigma\d\Sigma
\left[N_{\alpha\perp}\left(-\frac{\hbar^2}{2}\fdiff{^2}{\varphi^2} 
+ \frac{1}{2}(\naSig\varphi_\Delta)^2 + 
\frac{1}{2}m^2\varphi_\Delta^2\right) + 
N_\parallel\cdot\naSig\varphi_\Delta\,\hbar\fdiff{}{\varphi}\right] 
W[\varphi_\Delta,V_\Delta]\right\} \\
&& {}+ O(|V-V_\Delta|) \\
&=& 
-\frac{1}{\hbar}\int_\Sigma\d\Sigma\left[N_\perp
\left(-\frac{\hbar^2}{2}\fdiff{^2}{\varphi^2} 
+ \frac{1}{2}(\naSig\varphi)^2 + \frac{1}{2}m^2\varphi^2\right) + 
N_\parallel\cdot\naSig\varphi\,\hbar\fdiff{}{\varphi}\right] 
W[\varphi,V] + O(|V-V_\Delta|)
\eean
In the $\delta\to 0$ limit, $|V-V_\Delta|$ goes to zero and one 
recovers the generalized Schr\"odinger equation for curved boundaries.

\section{Summary}
\label{Summary}

We have proposed an exact definition for a Euclidean free scalar
propagator $W[\varphi,V]$ which ``evolves" wavefunctionals of fields
along general spacetime domains $V$.  Our main result is a derivation
of the evolution equation 
\be
\label{mainresult}
\hbar L_N W[\varphi,V] = \left(-H_N[\varphi,\hbar\fdiff{}{\varphi},V] 
+ P_N[\varphi,\hbar\fdiff{}{\varphi},V]\right)W[\varphi,V]\,.
\ee
This equation describes how $W[\varphi,V]$ varies under infinitesimal
deformations of $V$ generated by a vector field $N$.  The
variation is given by the action of two operators: one is related to the
field Hamiltonian and arises from normal deformations of the boundary
$\Sigma = \pa V$.  The second operator results from tangential
deformations and generalizes the field momentum.

We showed also that the Hamilton  function of the classical 
system satisfies an analogous Hamilton-Jacobi equation 
\be
\label{analogousHJequation}
L_N S[\varphi,V] = H_N[\varphi,\fdiff{S}{\varphi},V] + 
P_N[\varphi,\fdiff{S}{\varphi},V]\,.
\ee
When the boundary $\Sigma$ consists of two infinite hyperplanes at 
fixed times, \eq{mainresult} and \eq{analogousHJequation} reduce to 
the standard Schr\"odinger and Hamilton-Jacobi equation in their 
Euclidean form.

The derivation of eq.\ \eq{mainresult} is based on assumptions which
we consider plausible, but are not proven.  Most importantly, we have
not shown that the proposed regularization of the propagator has a
well-defined continuum limit.  A description for converting the
Euclidean to a Lorentzian propagator is missing.  As described in
section \ref{General_Boundary_Approach}, we expect that an evolution
equation analogous to \eq{mainresult} holds also for Lorentzian
propagators.  We emphasize that such state evolution may, in general,
be non-unitary and nevertheless admit a physical interpretation.

\subsection*{Acknowledgements}
We thank Luisa Doplicher, Robert Oeckl, Daniele Oriti, Massimo
Testa and Thomas Thiemann for helpful discussions. This work was supported by the Daimler-Benz foundation and DAAD. We thank the physics department of
the University of Rome ``La Sapienza" for the hospitality.

\begin{appendix}
\section{Local Form of Hamilton-Jacobi and Schr\"odinger Equation}

In the text we have presented the generalized Hamilton-Jacobi and
Schr\"odinger equation in an integral form.  They can be also
expressed locally, and below we explain how the two representations
are related.  The local notation is used in \cite{Rovelli} and
\cite{CDORT}.

Both the Hamilton  function $S$ and the propagator $W$ 
depend on the volume $V$. The latter is enclosed by the boundary 
$\Sigma$. Consider a parametrization of $\Sigma$, i.e.\ a map 
$$x: P\to\Sigma\,,\:\: \tau\mapsto x(\tau)$$
from a $(d-1)$-dimensional manifold $P$ to $\Sigma$. Provided it is 
clear on which ``side'' of $\Sigma$ the volume lies, one can view $S$ 
and $W$ as functions of $\Sigma$, or equivalently, as functionals of 
the parametrizing map $\tau\mapsto x(\tau)$. The other variable of 
$S$ and $W$ is the boundary field $\varphi: \Sigma\to\bR$: we may 
replace it by its pull-back $\tilde{\varphi}$ to $P$, so that $S$ and 
$W$ are completely expressed in terms of quantities on the parameter 
manifold $P$:
\bean
&& \tilde{\varphi}(\tau) = \varphi(x(\tau))\,,\:\: \tau\in P\,, \\
&& S = S[\tilde{\varphi}(\tau),x(\tau)]\,,\quad W = 
W[\tilde{\varphi}(\tau),x(\tau)]\,.
\eean
In section \ref{Generalized_Hamilton_Jacobi_Equation}, we defined the 
deformation derivative $L_N$ which acts by infinitesimal 
diffeomorphisms and pull-forwards of $V$ and $\varphi$ respectively. 
A moment's reflection shows that in the new notation the same effect 
is achieved by applying a variation 
$$\delta x(\tau) = s N(x(\tau))$$
to the function $x(\tau)$ while leaving $\tilde{\varphi}(\tau)$ 
fixed. Therefore,
\be
\label{LNequal}
L_N \equiv \int_P \d^{d-1}\tau\,N^\mu(x(\tau))\fdiff{}{x^\mu(\tau)}\,.
\ee
Our explicit result for the Hamilton-Jacobi equation (p.\ 
\pageref{generalizedHJ}, eq.\ \eq{generalizedHJ}) can be rewritten as
\bea
L_N S[\varphi,V] &=&
\int_P 
\d^{d-1}\tau\,N^\mu(x(\tau))\Bigg\{n_\mu(x(\tau))\left[-\frac{1}{2}\left(\fdiff{S}{\varphi(\tau)}\right)^2 
+ \frac{1}{2}(\nabla\phi(\tau))^2 + \frac{1}{2}m^2\phi^2(\tau) + 
U(\phi(\tau))\right] \nonumber \\
&&\hspace{3.5cm} 
{}-\nabla_{\!\mu}\varphi(\tau)\fdiff{S}{\varphi(\tau)}\Bigg\}\,, 
\eea
where on the right-hand side $S$ is a functional of the new 
variables. Comparison with \eq{LNequal} gives the equation
\bea
\label{localHJequation}
\fdiff{S}{x^\mu(\tau)} = 
n_\mu(x(\tau))\left[-\frac{1}{2}\left(\fdiff{S}{\varphi(\tau)}\right)^2 
+ \frac{1}{2}(\nabla\phi(\tau))^2 + \frac{1}{2}m^2\phi^2(\tau) + 
U(\phi(\tau))\right] - 
\nabla_{\!\mu}\varphi(\tau)\fdiff{S}{\varphi(\tau)}\,.
\eea
It describes how $S$ behaves under local variations of the boundary 
$\Sigma$. By the same reasoning, we arrive at a local Schr\"odinger 
equation for the kernel $W$:
\be
\label{localSchroedingerequation}
\fdiff{W}{x^\mu(\tau)} = 
n_\mu(x(\tau))\left[-\frac{\hbar^2}{2}\fdiff{^2W}{\varphi(\tau)^2} + 
\frac{1}{2}(\nabla\phi(\tau))^2 + \frac{1}{2}m^2\phi^2(\tau)\right] - 
\hbar\nabla_{\!\mu}\varphi(\tau)\fdiff{W}{\varphi(\tau)}\,.
\ee

\end{appendix}

\end{document}